\documentclass[12pt]{article}
\usepackage{theorem,amssymb,amsbsy,latexsym}

\newtheorem{theorem}{Theorem}[subsection]
 
\newtheorem{hypothesis}[theorem]{Hypothesis}
\newtheorem{lemma}[theorem]{Lemma}
\newtheorem{prop}[theorem]{Proposition}
\newtheorem{corl}[theorem]{Corollary}
\newtheorem{remark}[theorem]{Remark}

\newcommand{\Ref}[1]{(\ref{#1})}
\newcommand{\define}{\stackrel{\mbox{{\tiny def}}}{=}}
\newcommand{\mbf}[1]{{\boldsymbol {#1} }}

\newcommand{\QED}{\hfill$\square$}
\newcommand{\nonu}{\nonumber \\ \nopagebreak}
\newcommand{\sign}{{\rm sign}}
\newcommand{\ee}[1]{{\rm e}^{#1}}
\newcommand{\ii}{{\rm i}}
\newcommand{\dd}{{\rm d}}

\newcommand{\tet}{\theta}
\newcommand{\eps}{\varepsilon}
\newcommand{\half}{\mbox{$\frac{1}{2}$}}

\newcommand{\vx}{{\bf x}}
\newcommand{\vy}{{\bf y}}
\newcommand{\vz}{{\bf z}}
\newcommand{\vn}{{\bf n}}
\newcommand{\vlambda}{{\mbf \lambda}}

\newcommand{\vm}{{\bf m}}
\newcommand{\ve}{{\bf e}}
\newcommand{\vE}{{\bf E}}

\newcommand{\vnu}{\hat{\mbf \nu}}
\newcommand{\vzero}{{\mbf 0}}

\newcommand{\R}{{\mathbb R}}
\newcommand{\C}{{\mathbb C}}
\newcommand{\Z}{{\mathbb Z}}
\newcommand{\N}{{\mathbb N}}

\newcommand{\cO}{{\mathcal  O}}

\newcommand{\cE}{{\mathcal  E}}
\newcommand{\cC}{{\mathcal  C}}
\newcommand{\cJ}{{\mathcal  J}}

\newcommand{\opS}{{\mathbb S}} 
\newcommand{\opA}{{\mathbb A}}
\newcommand{\opB}{{\mathbb B}} 
\newcommand{\opP}{{\mathbb P}} 
\newcommand{\opPperp}{{\mathbb P}^{\perp}}

\newcommand{\newsection}{\setcounter{equation}{0}\section}

\renewcommand{\appendix}{\setcounter{equation}{0}\setcounter{section}{0}\renewcommand{\thesection}{\Alph{section}}}
\newcommand{\appsection}[1]{\setcounter{equation}{0}\renewcommand{\thesection}{\Alph{section}}
\section{#1} \renewcommand{\thesection}{\Alph{section}}}
\begin{document}

\begin{flushright}
December 8, 2008
\end{flushright}
\vspace{.4cm}

\begin{center}

{\Large \bf Explicit solution of the (quantum) elliptic
Calogero--Sutherland model}
\vspace{1 cm}

{\large Edwin Langmann}\footnote{langmann@kth.se}\\
\vspace{0.3 cm} 

\textit{Theoretical Physics, AlbaNova, SE-106 91 Stockholm, Sweden}

\bigskip

\textit{Dedicated to the memory of Vadim Kuznetsov}

\end{center}

\begin{abstract}
  We derive explicit formulas for the eigenfunctions and eigenvalues
  of the elliptic Calogero-Sutherland model as infinite series, to all
  orders and for arbitrary particle numbers and coupling parameters.
  The eigenfunctions obtained provide an elliptic deformation of the
  Jack polynomials. We prove in certain special cases that these
  series have a finite radius of convergence in the nome $q$ of the
  elliptic functions, including the two particle (= Lam\'e) case for
  non-integer coupling parameters.  \bigskip

\noindent {PACS: 02.30.Ik, 03.65.-w, 05.30.Pr\\ MSC-class: 81Q05,
35Q58}
\end{abstract}

\newsection{Introduction}
\label{sectINTRO}
In this paper we present an explicit solution of the elliptic
generalization of a quantum many body model in one dimension which is
usually associated with the names of Calogero and Sutherland
\cite{C1,Su,C2} and whose quantum integrability was proved by
Olshanetsky and Perelomov \cite{OP}. Our solution is based on a
remarkable functional identity stated in Lemma~\ref{fact1} below. This
identity was found in \cite{EL1,EL2} using quantum field theory
techniques \cite{CL}, but to make the present paper self-contained we
also include an alternative, elementary proof of this key result.  Our
main results are explicit formulas for the eigenvalues and
eigenfunctions of this model as infinite series. We also prove
absolute convergence of these series in certain special cases.

\subsection{Background} 
The elliptic Calogero-Sutherland (eCS) model is defined by the
differential operator
\begin{equation}
\label{eCS}
H \define - \sum_{j=1}^N\frac{\partial^2}{\partial x_j^2} \; + \; \gamma \!\!
\sum_{1\leq j<k\leq N} V(x_j-x_k) 
\end{equation}
with $-\pi\leq x_j \leq\pi$ coordinates on the circle, $N=2,3,\ldots$,
$\gamma>-1/2$, and the function
\begin{equation}
\label{V0}
V(r) \define \sum_{m\in\Z}\frac{1}{ 4\sin^2[(r+ \ii\beta m )/2]}
= \wp(r \,  | \, \pi, \ii\beta/2)+c_0 
\:  \quad
(\beta>0)    
\end{equation}
with the Weierstrass elliptic function $\wp$ and the constant  
\begin{equation}
\label{c0}
c_0 = \frac{1}{12} -\sum_{m=1}^\infty \frac{1}{2\sinh^2[(\beta m)/2]}
=  \frac{1}{12} -
\sum_{n=1}^\infty \frac{2 q^{2n}}{(1-q^{2n})^2} 
\end{equation} 
where 
\begin{equation}
q\define \ee{-\beta/2} 
\end{equation} 
(see Appendix~\ref{appA1} for more details); we find it convenient to
shift $\wp$ by a constant to simplify some formulas later on. We also
introduce the function
\begin{equation} 
\label{tet}
\tet(r) \define  \sin(r/2) \prod_{n=1}^\infty
(1-2q^{2n}\cos(r) + q^{4n})
\end{equation} 
which is equal, up to a multiplicative constant, to the Jacobi Theta
function $\vartheta_1(r/2)$. This allows us to write
\begin{equation}
\label{V}
V(r) \, = \, -\frac{d^2}{d r^2} \log \tet(r)
\end{equation}
(see Appendix~\ref{appA1} for details). This differential operator $H$
defines a quantum mechanical model of $N$ identical particles moving
on a circle of length $2\pi$ and interacting with a two body potential
$\gamma V(r)$ with $\gamma$ the coupling constant.  To be
more precise: the model we are interested in corresponds to a
particularly `nice' self-adjoint extension of this differential
operator \cite{KT} (which is essentially the Friedrichs extension
\cite{RS2}), and we only consider eigenfunctions describing
non-distinguishable particles. It is convenient to parametrize the
coupling constant as follows,
\begin{equation}
\label{gamma} 
\gamma \define 2\lambda(\lambda-1)  
\end{equation}
where $\lambda>1/2$ is the coupling parameter. Our restrictions on the
parameters $q$ and $\lambda$ are for simplicity and since they include
most of the cases where $H$ defines a self-adjoint operator bounded
from below, but many of our results can be analytically continued to
other complex parameter values.

In the limiting case $q=0$ ($\beta\to\infty$) we have
$V(r)=(1/4)\sin^{-2}(r/2)$, and the differential operator $H$ in
\Ref{eCS} reduces to the one defining a celebrated model solved a long
time ago by Sutherland \cite{Su} who found explicit formulas for all
eigenvalues and an algorithm to construct the corresponding
eigenfunctions of $H$. The exact eigenfunction $\Psi_\vn$ of the
Sutherland model can be labeled by \textit{partitions} $\vn$, i.e.\
integer vectors $\vn=(n_1,n_2,\ldots,n_N) \in \N_0^N$ such that
$n_1\geq n_2 \geq \ldots \geq n_N\geq 0$, and they are of the form
\begin{equation} 
\label{Psin} 
\Psi_\vn(\vx) =
\cJ_\vn(\vz) \Psi_0(\vx),\quad z_j\define \ee{\ii x_j}
\end{equation}
with 
\begin{equation} 
\label{Psi0}
\Psi_0(\vx) \define \prod_{1\leq j<k\leq N} \tet(x_k-x_j)^\lambda  
\end{equation}
and the $\cJ_\vn(\vz)$ symmetric polynomials with simple corresponding
eigenvalues given in \Ref{Enn} below.  The $\cJ_\vn$ are known as
\textit{Jack polynomials} and are also of interest in combinatorics;
see \cite{McD,St}.

It is interesting to note that, to get all eigenfunctions of interest
in physics, one should multiply the r.h.s.\ of \Ref{Psin} by a factor
$\ee{-\ii P(x_1+\cdots + x_N)}$ with an arbitrary $P\in \N$
corresponding to an additional possible center-of-mass motion.  One
can account for this by extending the definition of the Jack
polynomials to all integer vectors $\vn$ such that $n_1\geq n_2\geq
\cdots \geq n_N$ (with $n_N$ possibly negative) as follows,
\begin{eqnarray*}
  \cJ_{\vlambda -P\ve}(\vz) = c_{\vlambda,P}(z_1z_2\cdots z_N)^{-P} 
  \cJ_\vlambda(\vz) \:,  \quad \ve\define (1,1,\cdots,1)
\end{eqnarray*} 
for all partitions $\vlambda$ and $P\in\N$; the $c_{\vlambda,P}$ are
some non-zero constants depending on the normalization of the
$\cJ_\vn$.  This equation actually holds true also for negative
integers $P$; see e.g.\ \cite{St}. It is common to ignore this point
and restrict the $\vn$ to partitions.

Sutherland's solution method is based on the fact that, for $q=0$, the
differential operator $H$ in \Ref{eCS} has an exact eigenstate
$\Psi_0(\vx)$ of a form as in \Ref{Psi0}.  This is no longer true for
$q>0$ \cite{Su2}, and thus Sutherland's approach cannot be generalized
to the elliptic case. In this paper we elaborate another algorithm
which allows to solve also the general elliptic case \cite{EL1,EL5}.
In the trigonometric limit $q=0$, this algorithm simplifies to one
which is different from Sutherland's; see \cite{EL3} for a comparison
of these algorithms. The crucial difference is that, while Sutherland
computes the Jack polynomials $\cJ_\vn(\vz)$ as linear combinations of
the following basis in the space of symmetric polynomials,
$M_{\vlambda}(\vz) = \sum_{\pi} \prod_{j=1}^N
z_{j}^{\lambda_{\pi(j)}}$ for partitions $\vlambda$ and the sum over
all distinct permutations $\pi$ in the permutation group $S_N$, we use
a different generating set of functions $f_\vn(\vz)$; see \Ref{fn} for
$\Theta(z)=(1-z)$.  It is important to note that the $f_\vn$ are
well-defined and non-zero also for integer vectors $\vn$ which are not
partitions. Thus the $f_\vn$ provide an overcomplete generating set in
the space of symmetric polynomials. Expanding the $\cJ_\vn$ in this
generating set
\begin{equation}
\label{cJcP} 
\cJ_\vn(\vz) = \sum_{\vm} \alpha_\vn(\vm) f_\vm(\vz) \: ,
\end{equation}
the expansion coefficients $\alpha_\vn(\vm)$ obey simple
recursion relations \cite{EL3} which can be solved explicitly
\cite{EL5}. 

We emphasize that, in our approach, we get eigenfunctions $\Psi_\vn$
for all integer vectors $\vn$. Obviously, if $\cE_{\vn}$ for a
non-partition $\vn\in\Z^N$ is different from all $\cE_{\vlambda-P\ve}$
for partitions $\vlambda$ and $P\in\N$, then the corresponding
eigenfunction $\Psi_{\vn}$ must vanish, otherwise it must be a linear
combination of the $\Psi_{\vlambda-P\ve}$ such that
$\cE_{\vn}=\cE_{\vlambda-P\ve}$.\footnote{We have checked this for
  many different integer vectors $\vn$ and parameters $\lambda$ in the
  case $q=0$ and $N=2$ using MAPLE.} It would be interesting to
investigate this overcompleteness.

\subsection{The nature of our solution} 
For $q>0$, our algorithm leads to eigenfunctions $\Psi_\vn$ as in
\Ref{Psin}--\Ref{cJcP} with a natural generalization of the functions
$f_\vn(\vz)$ given in Proposition~\ref{prop1} below.  It is natural to
regard the functions $\cJ_\vn(\vz)$ thus defined as an elliptic
generalization of the Jack polynomials. It is important to note that
the latter are no longer polynomials for $q>0$; see Remark~1 in
Section~\ref{sec5} for a more precise characterization. Moreover, the
corresponding eigenvalues $\cE_\vn$ and the coefficients
$\alpha_\vn(\vm)$ are more complicated \cite{EL1}.

One main result in this paper is an implicit equation determining the
eigenvalues $\cE_\vn$ and an explicit formula for the coefficients
$\alpha_\vn(\vm)$ depending on $\cE_\vn$ (Theorem~\ref{thm1}). We also
present a solution of these equations by some variant of perturbation
theory to all orders. This leads to fully explicit formulas for
$\cE_\vn$ and $\alpha_\vn(\vm)$ as infinite series
(Theorem~\ref{thm2}). Our series are (essentially) power series in the
parameter $\gamma$, but it is important to note that the small
parameter is still $q^2$ and $\gamma$ is only a convenient book
keeping device. A simple analogue illustrating this latter feature of
our solution is the following function,
\begin{eqnarray*} 
f(q) = \sum_{s=2}^\infty \gamma^{s} \sum_{\nu_1,\ldots,\nu_s\in\Z}
\delta(\nu_1+\nu_2+\cdots + \nu_{s},0) \prod_{\ell=1}^{s}
\frac{|\nu_\ell| q^{2|\nu_\ell|}}{(1-q^{2|\nu_\ell|})}
\end{eqnarray*}
with $\delta$ the Kronecker delta: this function has a complicated
power series in $q^2$ while its power series $f(q)=\sum_{s=2}^\infty
\gamma^s f_s(q^2)$ in $\gamma$ is simple and can be written
explicitly. Moreover, it is not difficult to prove that
$f_s=\cO(q^{2\lceil\frac{s}{2}\rceil})$ and that this series has a
finite radius of convergence in $q^2$.

We also prove that our series for $\cE_\vn$ and $\alpha_\vn(\vm)$ have
a finite radius of convergence in $q^2$ under a certain hypothesis
which is fulfilled in several interesting cases, including $N=2$ for
non-integer $\lambda$ (Lam\'e case), and the groundstate $\vn=\vzero$
for all $N>2$ and irrational $\lambda$. Our estimates to prove
convergences are not optimal, and we believe that our series converge
for all possible parameter values. In numerical computations it might
be more efficient to solve the implicit equation for $\cE_\vn$ by
iteration rather than using our explicit series.

\subsection{Previous results} 
We now discuss previous results related to our work.

For $N=2$ the eigenvalue equation of the eCS differential operator can
be reduced to the so-called \textit{Lam\'e equation} which was studied
extensively at the end of the 19$^{th}$ century (see \cite{WW} for a
summary of these classical results) and more recently in
\cite{EK,R2}. It is known that, for integer values of $\lambda$, the
Lam\'e equation has a finite number (depending on $\lambda$) of
eigenfunctions and corresponding eigenvalues which can be computed
algebraically \cite{WW}, and a generalization of this result to $N>2$
was found by G\'omez-Ullate \textit{et.al} \cite{GGR} and led to
particular many-body generalizations of the Lam\'e equation which,
however, is different from the eCS model. There exist various other
interesting results on the solution of the eCS model for integer
values of $\lambda$: Dittrich and Inozemtsev obtained explicit
formulas for the eigenvalues of the eCS model for the cases
$\lambda=2$ and $N=3$ \cite{DI,I1} and $\lambda=2$ and general $N$
\cite{I2}.  A Bethe ansatz solution of the integer-$\lambda$ eCS
models and, more generally, the elliptic Ruijsenaars model \cite{R1}
(which reduces to the eCS model in a certain limit), was given by
Felder and Varchenko \cite{FV2,FV3}; see also \cite{B,HSY,T}. There
exists an interesting separation-of-variable approach to the solution
of the $N=3$ eCS model by Sklyanin \cite{S}.  The latter approach
seems to be similar to ours in that it is also based on an identity
closely related to the one in Lemma~\ref{fact1}, and it was elaborated
in detail in the trigonometric limit $q=0$ by Kuznetsov
\textit{et.al.}  \cite{KMS}.  We also mention the work of
N\'{u}\~{n}ez \textit{et.al} \cite{P} who computed the eigenvalues of
the eCS model for arbitrary values of $\lambda$ up to $\cO(q^2)$ for
$N\leq 4$ and up to $\cO(q^4)$ for $N=2$ and which our results extend
to arbitrary orders and particle numbers. Other related work is by
Komori and Takemura \cite{KT} who presented a perturbative algorithm
to solve the eCS model and proved that Schr\"odinger perturbation
theory has a finite radius of convergence in $q^2$; see also \cite{T}.
As far as we can see is the approach in \cite{KT} different from ours
(we obtain results which are different from standard ones even for
$q=0$ \cite{EL3}) and has not yielded results equally explicit as
ours.  A complimentary discussion of known results about eCS type
systems and further references can be found in \cite{KR}.

As mentioned, our approach is based on a remarkable functional
identity in Lemma~\ref{fact1}. This identity can be regarded as a
natural quantum analogue of the B\"acklund transformation discovered
by Wojciechowski \cite{W}; see also \cite{KS}. Similar identities were
obtained by Felder and Varchenko \cite{FV1} and, for $N=3$, by
Sklyanin \cite{S}. Moreover, its trigonometric limit is equivalent to
a well-known identity for the Jack polynomials due to Stanley; see
Proposition~2.1 in Ref.\ \cite{St}. Similar identities exist for a
large class of Calogero-Sutherland type models and provide a tool to
construct explicit formulas for the eigenfunctions of all these
systems in a unified way \cite{HL}. It is interesting to note that the
functions $f_\vn(\vz)$ for $q=0$ are the building blocks of this
explicit solution for all these models.

The results presented in this paper evolved over several years. The
key to our solution was found already in 2000 \cite{EL1}, but we
realized only in 2004 that it is possible to obtain an explicit
solution to all orders \cite{EL5}. Readers interested in how this
result emerged can consult the two earlier versions of the present
paper \cite{EL4}.  The first version (\cite{EL4} v1) is a manuscript
from 2001 which remained unpublished until January 2004. It contains
our solution algorithm together with a recipe how to compute the
eigenvalues as power series in the nome $q$ of the elliptic
functions. In the second version (\cite{EL4} v2) published in April
2004 we added explicit formulas for the eigenvalues up to $\cO(q^8)$
for $N=2$ and up to $\cO(q^4)$ for $N\leq 4$. These formulas were
obtained by straightforward but tedious computations. These latter
results suggested an alternative, more efficient, method allowing to
find the solution to all orders in $q^2$. This result was previously
announced in Refs.\ \cite{EL5,EL7} and is elaborated and extended in
the present paper.

A key issue in this evolution of our result was that we realized that
there exist better answers to the question: ``What parameter should we
use to expand our solution in?'', than the obvious one.  The obvious
answer is: ``$q$'' \cite{EL1}. However, this leads to very complicated
formulas already at low orders. A better answer is: ``$\gamma$'',
since it leads to much simpler formulas at low order; see
Section~\ref{sec3}.  The final answer which allowed us to deduce
explicit formulas to all orders is: ``$\eta$'', where $\eta$ an
auxiliary parameter which we introduce into the problem and which can
be set to $1$ at the end of the computation. In particular, the
explicit formulas for the eigenvalues which we thus obtain are of the
form $\cE_\vn=\cE_0(\vn)+\sum_{m=1}^\infty \eta^m\cE_m(\vn)$ where
$\cE_m(\vn)=\cO(q^{2m})$ are Taylor series in $q^2$; see
Theorem~\ref{thm2} and Remark~\ref{rem5}.

\subsection{Plan of the rest of the paper} 
In the next section we summarize preliminary results which we need. In
Section~\ref{sec3} we derive the solution of the eCS model, i.e.\ its
eigenfunctions and corresponding eigenvalues, as a series in $\gamma$
up to order $\cO(\gamma^3)$ and $\cO(\gamma^4)$, respectively, using
an elementary argument.  Our main results are obtained in
Section~\ref{sec4}: explicit formulas for the solution of the eCS
model as a series in $\gamma$ to all orders, together with a
sufficient condition for absolute convergence (Theorems~\ref{thm1} and
\ref{thm2}). We end with remarks in Section~\ref{sec5}. Details of our
computations and some proofs are deferred to three appendices.

\noindent \textbf{Notation:} We denote as $\Z$, $\Z'$, $\N_0$ and $\N$
the sets of all, all non-zero, all non-negative, and all positive
integers, and $\C$ and $\R$ are the complex and real numbers,
respectively. We use bold symbols for vectors with $N$ components,
e.g.\ $\vn\in\Z^N$ is short for $(n_1,\ldots,n_N)$ with $n_j\in\Z$
etc. The symbol ``$\delta$'' always means the Kronecker delta, in
particular, $\delta(\vm,\vn)=\prod_{j=1}^N\delta_{m_j,n_j}$ for
$\vm,\vn\in\Z^N$. For $x\in\R$ we denote as $\lceil x\rceil$ the
smallest integer larger or equal to $x$. Definitions are indicated by
the symbol ``$\define$''.

\newsection{Summary of preliminary results} 
\label{sec2}
In this Section we collect a few results which we need.

\subsection{A remarkable identity} 
The starting point for our solution method is the following.

\begin{lemma}
\label{fact1} 
Let
\begin{equation}
\label{F}
F(\vx;\vy) \define \frac{ \prod_{1\leq j<k\leq N} \tet(x_{k}-x_j)^{\lambda} 
  \prod_{1\leq j<k\leq N} 
  \tet(y_{j}-y_{k})^{\lambda}}{\prod_{j,k=1}^N
  \tet(x_j-y_k)^{\lambda}} \: ,   
\end{equation}
with $\tet(r)$ in \Ref{tet}, where $\vx = (x_1,\ldots,x_N)\in\C^N$ and
similarly for $\vy$. Then the following identity holds true,
\begin{equation}
\label{rem}
\sum_{j=1}^N\biggl(\frac{\partial^2}{\partial x_j^2} -  
\frac{\partial^2}{\partial y_j^2}  \biggr)F(\vx;\vy) 
=  2\lambda(\lambda-1)\sum_{1\leq j<k\leq N}\biggl( 
V(x_k-x_j) - V(y_j-y_k) \biggr)F(\vx;\vy)  
\end{equation}
with $V(r)$ as in \Ref{V}. 
\end{lemma}

\noindent (Proof in Appendix~\ref{appA3}.) 

As already mentioned, the result in Lemma~\ref{fact1} was obtain in
\cite{EL2} using quantum field theory techniques.  The proof given in
Appendix~\ref{appA3} is elementary and based on the following
functional identity \cite{WW}
\begin{equation}
\label{rel}
\phi(x)\phi(y) +\phi(x)\phi(z)+ \phi(y)\phi(z) 
= f(x)+f(y)+f(z)
\quad \; \mbox{ if $x+y+z=0$} 
\end{equation}
where 
\begin{equation}
\label{defrel}
\phi(x) = \frac{d}{dx}\log\tet(x)\, , \quad f(x) = \half
    [V(x)-\phi(x)^2 - c_0]
\end{equation}
and $c_0$ the constant in \Ref{c0}; a proof of the latter identity is
in Appendix~\ref{appA2}.

It is important to note that one can write the identity in \Ref{rem}
as follows,
\begin{equation}
\label{rem1}
H(\vx) F(\vx;\vy) = H(\vy) F(\vx;\vy)
\end{equation}
where $H$ is the differential operator in \Ref{eCS} but acting on
different arguments $\vx$ and $\vy$, as indicated.

\subsection{Reformulation of the eigenvalue problem} 
From Lemma~\ref{fact1} a straightforward computation leads to a result
which is the next step in our solution. To state this result we find
it convenient to define
\begin{equation} 
\label{Ejk}
(\vE_{jk})_\ell \define \delta_{j\ell}-\delta_{k\ell} \quad \forall j,k,\ell
= 1,\ldots, N,\quad j<k 
\end{equation}
and use the shorthand notation
\begin{equation}
\label{vnu}
\sum_{\vnu} S_\nu \alpha(\vnu) \define \sum_{1\leq j<k\leq N} 
\sum_{\nu\in\Z} S_\nu \alpha( \nu\vE_{jk})  
\end{equation}
for functions $\alpha$ on integer vectors in $\Z^N$.

\begin{prop}
\label{prop1} 
Let 
\begin{equation}
\label{Fhat}
\hat F_\vn(\vx) \define f_\vn(\vz)\Psi_0(\vx) \, , 
\quad \vn\in\Z^N \; \mbox{ and }\;  z_j=\ee{\ii x_j} 
\end{equation}
with $\Psi_0(\vx)$ defined in \Ref{Psi0} and \Ref{tet} and the special
functions
\begin{equation}
\label{fn}
f_\vn(\vz) \define \left( \prod_{j=1}^N \oint_{\cC_j}\frac{\dd \xi_j}{2\pi\ii
\xi_j}\xi_j^{n_j} \right) 
\frac{\prod_{1\leq j<k\leq N}\Theta(\xi_j/\xi_k)^\lambda
}{\prod_{j,k=1}^N\Theta(z_j/\xi_k)^\lambda }
\end{equation}
with the integration paths
\begin{equation} 
\label{cCj}
\cC_j:\: \xi_j = \ee{\eps_j}\ee{\ii \varphi_j}\, , \quad -\pi\leq
\varphi_j\leq \pi\, , \quad 0<\eps_1<\eps_2<\cdots < \eps_N <\beta
\end{equation}
and
\begin{equation}
\label{Theta} 
\Theta(z) \define (1-z) \prod_{m=1}^\infty\left[ \left(1-q^{2m} z
\right)\left(1-q^{2m}/z \right) \right] \: .
\end{equation}
Then the eCS differential operator $H$ in \Ref{eCS}--\Ref{V0} obeys
\begin{eqnarray}
\label{HFhat}
H \hat F_\vn(\vx) = \cE_0(\vn) \hat F_\vn(\vx) - \gamma \sum_{\vnu}
S_\nu \hat F_{\vn+\vnu}(\vx)
\end{eqnarray}
where
\begin{equation}
\label{Enn} 
\cE_0(\vn) \define \sum_{j=1}^N \left(n_j + \half\lambda(N+1-2j) \right)^2
\end{equation}
and 
\begin{equation} 
\label{Snu} 
S_\nu \define |\nu| \frac{q^{|\nu|-\nu}}{1-q^{2|\nu|}}\quad \forall
\nu\in\Z',\quad S_0\define 0\: .
\end{equation}
\end{prop}

\noindent (Proof in Appendix~\ref{appB}.)

This result concluded our discussion in \cite{EL2}.\footnote{Note that
  $f_\vn(\vz)$ here was denoted as ${\mathcal P}(\vn;\vx)$ in
  \cite{EL2}.} Our proof in Appendix~\ref{appB} is (essentially) by
expanding the identity in \Ref{rem1} in a Laurent series in the
variables $\xi_j=\ee{\ii y_j}$ in a certain region in $\C^N$ and
equating the expansion coefficients.

As discussed below, the integrals in \Ref{fn}--\Ref{Theta} are
well-defined and independent of the parameters $\eps_j$ in the
specified range due to Cauchy's theorem. It is important to note that
$S_\nu=0$ for $\nu\leq 0$ and $q=0$. As we will see, this simplifies
the solution for $q=0$ drastically.

An important consequence of Proposition~\ref{prop1} is the following.

\begin{corl} \label{corl1} 
Let 
\begin{equation}
\label{ansatz}
\Psi(\vx) = \sum_{\vm\in\Z^N} \alpha(\vm) \hat F_\vm(\vx)
\end{equation}
with coefficients $\alpha(\vm)$ satisfying the following relations,
\begin{equation} 
\label{eq}
[\cE_0(\vm) -\cE ] \alpha(\vm) = \gamma (\opS\alpha)(\vm)
\end{equation}
for some constant $\cE$ and
\begin{equation}
\label{opS} 
(\opS\alpha)(\vm) \define \sum_{\vnu} S_\nu \alpha(\vm-\vnu)
\end{equation}
with $\hat F_\vn(\vx)$, $\cE_0(\vn)$, and $S_\nu$ defined in
Proposition~\ref{prop1} and $\vnu$ in \Ref{vnu}. Then $\Psi(\vx)$ is
an eigenfunction of the eCS Hamiltonian in \Ref{eCS} with eigenvalue
$\cE$: $H\Psi(\vx)=\cE\Psi(\vx)$.
\end{corl}

This is a simple consequence of the ansatz \Ref{ansatz} and
Proposition~\ref{prop1} which imply
\begin{eqnarray*}
(H-\cE)\Psi = \sum_{\vm}\Bigl( [\cE_0(\vm) -\cE]\alpha(\vm) -
\gamma(\opS\alpha)(\vm) \Bigr) \hat F_\vm(\vx) \: .
\end{eqnarray*}
Thus the problem of solving the eCS model is reduced to finding
solutions $\cE$ and $\alpha(\vm)$ of \Ref{eq}.

\subsection{Properties of the functions $f_\vn(\vz)$} 
The functions $f_\vn(\vz)$ defined in \Ref{fn}--\Ref{Theta} play an
important role in our solution. To see that they are well-defined we
note that
\begin{equation} 
\label{Thetabound} 
0<\Theta(|z|)\leq |\Theta(z)|\leq \Theta(-|z|)<\infty \;
\mbox{ for }\; q^2 < |z| < 1 
\end{equation}
(see Appendix~\ref{cPbound}), and thus the integrand in \Ref{fn} is
analytic in the region $1 <|\xi_1|<|\xi_2|<\cdots<|\xi_N|< q^{-2}$ if
all $|z_j|=1$. Thus the integral in \Ref{fn} are well-defined and
independent of the closed integration paths $\cC_j$ as long as they
are within this region of analyticity (Cauchy's theorem), which is
obviously the case for the ones in \Ref{cCj}.

It is important to note that the $f_\vn(\vz)$ are symmetric functions:
$f_\vn(z_1,z_2,\ldots,z_N) =
f_\vn(z_{\pi(1)},z_{\pi(2)},\ldots,z_{\pi(N)})$ for all $\pi\in
S_N$. Moreover, they are polynomials for $q=0$ \cite{EL3}. As already
mentioned, for $q=0$ the $f_\vn(\vz)$ provide a basis in the space of
symmetric polynomials if the $\vn$ are restricted to partitions, even
though they are, in general, non-zero even for non-partition integer
vectors $\vn$ \cite{EL3,HL}. We expect that similar results hold true
also for $q\neq 0$.

We finally state an upper bound for the functions $f_\vn(\vz)$ which
we need to prove square integrability of our eigenfunctions:

\begin{lemma}\label{lemPn}
  The functions $f_\vn(\vz)$ defined in \Ref{fn} obey
\begin{equation} 
|f_\vn(\vz)| < C \, q^{\sum_j ( \tilde{K}|n_j| - K j n_j )}\: , \quad
 |z_j|=1 
\end{equation}
with the constants
\begin{equation} 
  K\define \frac{2}{N+b},\quad \tilde{K}\define \frac{bK}{1+2b}, \quad C 
  \define
  \frac{[2\Theta(-q^2)]^{N(N-1)\lambda/2}}{ [
    (1-q^{K-\tilde{K}})\Theta(q^{2-2b\tilde{K}})/ (1-q^{2-2b\tilde{K}})]^{N^2\lambda}}  
\label{KC}
\end{equation}
for arbitrary $b>0$. 
\end{lemma}

\noindent (Proof in Appendix~\ref{cPbound}.)

Note that $0<\tilde{K}<K/2$, $2-2b\tilde{K}>0$, and $C<\infty$. It is
worth mentioning that these estimates are not optimal (and in fact
meaningless for $q=0$), but they are good enough for our purposes.

\subsection{Lagrange's reversion theorem}
\label{sec2.4} 
We finally recall a well-known result due to Lagrange which will play
a key role in our solution:

\begin{theorem} 
  \label{lagthm} (A) Let $\varphi(z)$ and $g(z)$ be functions of the
  complex variable $z$ analytic on and inside a closed contour $\cC$
  surrounding a point $z=a$, and $\eta$ a complex number such that
\begin{equation}
\label{tphi-condition}
\sup_{z\in\cC} \frac{|\eta\varphi(z)-a|}{|z-a|} <1\: .
\end{equation}
Then the equation $z = \eta\varphi(z)$ has a unique solution $z=\xi$
inside of $\cC$, $g(\xi)$ can be expanded as
\begin{equation} 
\label{Lagrange1} 
g(\xi) = g(a) + \sum_{m=1}^\infty \frac{\eta^m}{m!}
\frac{\dd^{m-1}}{\dd a^{m-1}} \left(\varphi(a)^m\frac{\dd}{\dd a}
  g(a) \right)\: , 
\end{equation}
and this series is absolutely convergent.

\noindent (B) The result in \Ref{Lagrange1} holds true in the sense of
formal power series even for values of $\eta$ where
\Ref{tphi-condition} is not fulfilled.
\end{theorem}

\noindent \textit{Proof:} Part (A) is equivalent to Lagrange's theorem
as stated in \cite{WW}, Section 7.32. The result in (B) is a simple
implication.  \QED

\begin{remark}
\label{remLagrange} 
We note that \Ref{Lagrange1} does not rely on analyticity and applies
to even more general situations where one knows that $\varphi(z)$ and
$g(z)$ have formal power series in $(z-a)$ but has no information
about convergence of these series. Indeed, inserting
\begin{equation}
\label{phi-series} 
\varphi(z)=\sum_{n=0}^\infty\varphi_n (z-a)^n
\end{equation} 
into the equation $z=\eta\varphi(z)$ and making an ansatz
$\xi=a+\sum_{n=1}^\infty \xi_k\eta^k$ for its solution one can solve
for the $\xi_k$ recursively and compute
\begin{equation}
\label{xi33}
\xi = a + \eta \varphi_0 + \eta^2 \varphi_0\varphi_1 + 
\eta^3 (\varphi_0^2\varphi_2 + \varphi_0\varphi_1^2) + 
\eta^4 (\varphi_0^3\varphi_3 + 3\varphi_0^2\varphi_1\varphi_2 +
\varphi_0\varphi_1^3) + \ldots \: . 
\end{equation}
Inserting this in $g(z)=\sum_{n=0}^\infty g_n (z-a)^n$ yields
\begin{eqnarray} 
\label{gseries}
g(\xi) & = &g_0 + \eta g_1\varphi_0 + \eta^2 (g_2 \varphi_0^2 + g_1
\varphi_0\varphi_1) + \eta^3 (g_3 \varphi_0^3 + 2g_2
\varphi_0^2\varphi_1 \nonu &+& g_1 [\varphi_0^2\varphi_2 +
\varphi_0\varphi_1^2]) + \eta^4 (g_4 \varphi_0^4 +
3g_3\varphi_0^3\varphi_1 + g_2 [2\varphi_0^3\varphi_2 + 3
\varphi_0^2 \varphi_1^2] \nonu &+& g_1 [\varphi_0^3\varphi_3
+3\varphi_0^2\varphi_1\varphi_2 +\varphi_0\varphi_1^3] ) + 
\ldots  
\end{eqnarray}
which can be straightforwardly extended to higher powers in $\eta$ and
agrees with \Ref{Lagrange1} to all orders in $\eta$.
\end{remark}

\newsection{Perturbative solution}
\label{sec3}
We will present our solution to all orders in the next section. As a
heuristic motivation, we present in this section a pedestrian approach
allowing to compute the first few terms of our series solution in a
simple manner.

We found that \Ref{eq} can be efficiently solved by making the ansatz
\begin{equation}
\label{s-ansatz} 
\alpha(\vm) = \sum_{s=0}^\infty \gamma^s \alpha^{(s)}(\vm),\quad \cE =
\sum_{s=0}^\infty \gamma^s \cE^{(s)}
\end{equation}
leading to the following system of equations equivalent to \Ref{eq},
\begin{equation} 
\label{IC} 
[\cE_0(\vm) -\cE^{(0)}] \alpha^{(0)}(\vm) = 0 
\end{equation}
and
\begin{equation} 
\label{recurr} 
[\cE_0(\vm) -\cE^{(0)}] \alpha^{(s)}(\vm) = \sum_{s'=1}^s \cE^{(s')}
\alpha^{(s-s')}(\vm) + (\opS\alpha^{(s-1)})(\vm) \quad \forall s>0 \:
.
\end{equation}
As mentioned in the introduction, $\gamma$ here is \textit{not}
assumed to be small but only serves as a convenient book keeping
parameter to organize our solution. It is important to note that, for
each $\vn\in\Z^N$, the initial condition in \Ref{IC} has the following
solution,
\begin{equation} 
\label{IC1} 
\cE^{(0)}_{\vn}=\cE_0(\vn),\quad \alpha^{(0)}_{\vn}(\vm) =
\delta^{\phantom 0}_\vn(\vm)
\end{equation}
where $\delta_\vn(\vm) \define \delta(\vm,\vn)$. Note that
$\alpha_{\vn}^{(s>0)}(\vn)$ is then undetermined by \Ref{recurr} and
could be set to any value. This ambiguity corresponds to the freedom
of multiplying the eigenfunction by an $\gamma$-dependent
constant. Our choice below is a convenient normalization.

Equation \Ref{recurr} implies
\begin{equation}
\label{cEs} 
 \cE_{\vn}^{(s)} = - (\opS\alpha_{\vn}^{(s-1)})(\vn) \quad \forall s>0
\end{equation}
where we set $\alpha_{\vn}^{(s)}(\vn)=0$ for all $s>0$, and
\begin{equation}
\label{als} 
\alpha_{\vn}^{(s)}(\vm) =
\frac1{b_\vn(\vm-\vn)}\Bigl(\sum_{s'=1}^{s-1} \cE_{\vn}^{(s')}
\alpha_{\vn}^{(s-s')}(\vm) + (\opS\alpha_{\vn}^{(s-1)})(\vm) \Bigr) \quad
\forall s>0 \: ; 
\end{equation}
we use the convenient shorthand notation
\begin{equation} 
\label{bnm}
\frac1{b_\vn(\vm-\vn)} \define \left\{ \begin{array}{ll} 
0 & \mbox{ if }\; \vm=\vn \\ 
{[}\cE_0(\vm) -\cE_0(\vn){]}^{-1} & \mbox{ otherwise } 
\end{array} \right. \; . 
\end{equation}

We thus can compute, recursively, all $\cE_{\vn}^{(s)}$ and
$\alpha_{\vn}^{(s)}(\vm)$. We obtain
\begin{eqnarray} 
\label{alpha3}
\cE_\vn^{(1)} = -\sum_{\vnu\in\Z} S_\nu \delta(\vzero,\vnu) = 0,
\nonu
\alpha_\vn^{(1)}(\vm) = \sum_{\vnu\in\Z} S_\nu
\frac{\delta(\vm-\vn,\vnu)}{b_\vn(\vnu)}, 
\nonu
\cE_\vn^{(2)} = -\sum_{\vnu_1,\vnu_2}
S_{\nu_1}S_{\nu_2} \frac{\delta(\vzero,\vnu_1+\vnu_2)}
{b_\vn(\vnu_1)} = - \sum_{\vnu\in\Z}
S_{\nu}S_{-\nu} \frac1{b_\vn(\vnu)}, 
\nonu
\alpha_\vn^{(2)}(\vm) = \sum_{\vnu_1,\vnu_2} S_{\nu_1}S_{\nu_2}
\frac{\delta(\vm-\vn,\vnu_1+\vnu_2)} {b_\vn(\vnu_1+\vnu_2)
b_\vn(\vnu_1)}, 
\nonu
\cE_\vn^{(3)} = -
 \sum_{\vnu_1,\vnu_2,\vnu_3}S_{\nu_1}S_{\nu_2}S_{\nu_3}
\frac{\delta(\vzero,\vnu_1+\vnu_2+\vnu_3)}
{b_\vn(\vnu_1+\vnu_2)b_\vn(\vnu_1)} , 
\nonu
\alpha_\vn^{(3)}(\vm) =
\sum_{\vnu_1,\vnu_2,\vnu_3}S_{\nu_1}S_{\nu_2}S_{\nu_3}\Bigl(
-\frac{\delta(\vzero,\vnu_1+\vnu_2)\delta(\vm-\vn,\vnu_3)}
{b_\vn(\vnu_1)b_\vn(\vnu_3)^2} + \nonu
\frac{\delta(\vm-\vn,\vnu_1+\vnu_2+\vnu_3)}
{b_\vn(\vnu_1+\vnu_2+\vnu_3 ) b_\vn(\vnu_1+\vnu_2)b_\vn(\vnu_1)}
\Bigr)
\nonu
\cE_\vn^{(4)} = 
 \sum_{\vnu_1,\vnu_2,\vnu_3}S_{\nu_1}S_{\nu_2}S_{\nu_3}S_{\nu_4} 
\Bigl(
\frac{\delta(\vzero,\vnu_1+\vnu_2)\delta(\vzero,\vnu_3+\vnu_4)}
{b_\vn(\vnu_1)b_\vn(\vnu_3)^2} - \nonu
\frac{\delta(\vzero,\vnu_1+\vnu_2+\vnu_3+\vnu_4)}
{b_\vn(\vnu_1+\vnu_2+\vnu_3 ) b_\vn(\vnu_1+\vnu_2)b_\vn(\vnu_1)}
\Bigr)
\end{eqnarray}
etc. It is straightforward but tedious to continue this computation to
higher orders in $\gamma$. Below we present a more efficient
computation method allowing us to derive closed formulas for $\cE_\vn$
and $\alpha_\vn(\vm)$ to all orders in $\gamma$.

It is straightforward to expand $\cE_\vn^{(s)}$ and $\alpha_\vn^{(s)}$
in powers of $q^2$, but the resulting formulas are rather complicated,
and we therefore do not present them here.\footnote{The interested
  reader can find a few such formulas \cite{EL4} v2.}

We thus find, for each $\vn\in\Z^N$, an eigenfunction $\Psi_\vn(\vx)$
and a corresponding eigenvalue $\cE_\vn$ as a formal power series in
$\gamma$. However, it is important to note the following potential
problem: It is possible that $\cE_0(\vm)-\cE_0(\vn)$ vanishes for some
$\vm \neq\vn$ with $\sum_{j}(m_j-n_j)=0$ which appear in the above
sums (recall that these series are built of terms $b_\vn(\vm-\vn)$ in
\Ref{bnm} with $\vm=\vn+\sum_{\kappa=1}^s\vnu_\kappa\neq \vn$, and all
such $\vm$ obey this latter condition), and in this case the sums are
not defined. We refer to such $\vm$ as \textit{resonances}. Thus the
results in the present section make sense only if there are no
resonances, i.e.\ if
\begin{equation}
\label{NoR}
\cE_0(\vm)\neq \cE_0(\vn) \quad \forall \vm\neq \vn\; \mbox{
  with }\;  \sum_{j=1}^N (m_j-n_j)=0 \: . 
\end{equation} 
Since there exist parameters $\lambda$ and $\vn$ where resonances
exist this is an important restriction.

We now discuss cases with resonances.  For example, for $N=2$ we have
\begin{equation}
\label{dE_N=2}
\cE_0(\vm)-\cE_0(\vn) = 2\nu(\nu+n_1-n_2+\lambda)\; , \quad
\nu\define m_1-n_1=-(m_2-n_2) \quad (N=2)\: ,
\end{equation}
and thus there is a resonance for $\nu=n_2-n_1-\lambda$ provided that
$\lambda$ is an integer. It is easy to see that there exist resonances
for all $\vn$ and $N$ if $\lambda$ is an integer, but for $N>2$ there
exist even resonances for non-integer $\lambda$. For example, for
$N=3$ and $\vn$ such that $\nu\define (n_1-2n_2+n_3)/3$ is a non-zero
integer, there is a resonance at $\vm=(n_1-\nu,n_2+2\nu,n_3-\nu)$ for
arbitrary values of $\lambda$.  Such $\lambda$-independent resonances
exist for all $N>2$.  It is also interesting to note that, if $N>2$,
there exist infinitely many ways to represent a resonance as
$\vm=\vn+\sum_{\kappa=1}^m\vnu_\kappa$ for $m>2$: this is true since
\begin{equation}
\label{cycle} 
\vE_{j_1j_2}+\vE_{j_2j_3} + \ldots +
\vE_{j_{l-1}j_l}-\vE_{j_1j_l}=\vzero\quad \;\mbox{ if }\;  
1\leq j_1<j_2<\cdots <
j_l \leq N,\quad l\leq N \: . 
\end{equation}
Due to this identity resonance can occur in the above sums infinitely
many times.

However, it is important to note that there exist cases without
resonances where the results of the present section apply. In
particular, \Ref{dE_N=2} shows that no resonance exists for $N=2$ and
non-integer $\lambda$. Moreover, results from numerical experiments
performed with MAPLE suggest that, for arbitrary particle numbers and
non-integer $\lambda$, there exist infinitely many $\vn$ without
resonances.

Anyway, we show in the next section that this potential problem of
resonances can be circumvented by resummations.

\newsection{Solution to all orders} 
\label{sec4}
We now introduce some notation which allows us to extend the solution
above to all orders.

\subsection{Implicit solution}
\label{sec4.1} 
We write \Ref{eq} as
\begin{equation} 
\label{eq3} 
(\opA -\cE) \alpha = \gamma \opS\alpha
\end{equation}
with
\begin{equation} 
(\opA\alpha)(\vm) \define \cE_0(\vm)\alpha(\vm) \: .  
\end{equation}
It is natural to interpret $\opA$ and $\opS$ as linear operators on
the vector space $\mathrm{Map}(\Z^N;\C)$ of functions $\alpha: \Z^N\to
\C$, $\vm\mapsto \alpha(\vm)$. (The topology of this space plays no
role in our discussion and is therefore ignored.) It is also
convenient to introduce the following projection on
$\mathrm{Map}(\Z^N;\C)$,
\begin{equation} 
(\opP_\vn\alpha)(\vm) \define \delta(\vm,\vn)\alpha(\vm) \: .  
\end{equation}
The equation in \Ref{eq3} can be solved using the following simple but
powerful result.

\begin{lemma}
\label{SolutionLemma}
Let $\opA$ and $\opB$ be linear operators on a vector space $V$ and
consider the eigenvalue equation
\begin{equation}
\label{AB}  
(\opA - \cE)\alpha = \opB \alpha 
\end{equation}
with the eigenvalue $\cE$. Let $\opP$ be some projection on $V$
commuting with $\opA$ and $\opPperp\define I-\opP$. Then
\begin{equation}
\label{AB1} 
\alpha = [I + (\opA - \cE - \opPperp \opB)^{-1}
\opPperp\opB] \alpha_0   
\end{equation}
is a solution of this eigenvalue equation provided that $\cE$ and
$\alpha_0$ satisfy the following conditions,
\begin{equation} 
\label{alpha0} 
\opP\alpha_0=\alpha_0
\end{equation}
and 
\begin{equation} 
\label{AB2} 
\cE\alpha_0 = \opA \alpha_0 - \opP\opB[I + (\opA - \cE - \opPperp
\opB)^{-1} \opPperp\opB] \alpha_0\: . 
\end{equation}
\end{lemma}

\noindent \textit{Proof:} Applying the projections $\opP$ to \Ref{AB}
and inserting
\begin{eqnarray*}
\alpha = \alpha_0 + \opPperp \alpha , \quad \alpha_0 \define \opP\alpha
\end{eqnarray*}
we obtain
\begin{equation}
\label{AB3} 
\cE \alpha_0 = \opA   \alpha_0 - \opP \opB \alpha 
\end{equation}
where we used that $\opA$ and $\opP$ commute. In a similar manner,
applying $\opPperp$ to \Ref{AB}, we get
\begin{eqnarray*} 
\opA \opPperp \alpha -  \cE \opPperp\alpha  - \opPperp \opB \opPperp
\alpha = \opPperp \opB \alpha_0  \: . 
\end{eqnarray*}
Obviously the last two equations together are equivalent to
\Ref{AB}. From the last equation we get
\begin{eqnarray*}
\opPperp \alpha = (\opA - \cE - \opPperp \opB)^{-1} \opPperp\opB
\alpha_0 \: ,
\end{eqnarray*} 
which implies \Ref{AB1}. Inserting \Ref{AB1} in \Ref{AB3} we obtain
the condition in \Ref{AB2}.  \QED

If we apply this result to \Ref{eq3} using the projection $\opP=
\opP_\vn$ and $\opB= \gamma\opS$ we can compute a solution $\alpha=
\alpha_\vn$ and $\cE=\cE_\vn$. It is easy to solve \Ref{alpha0}: its
general solution is $\alpha_0 = c \delta_\vn $ with
$\delta_\vn(\vm)=\delta(\vn,\vm)$ and an arbitrary constant $c$. It is
obvious that $c$ is an overall normalization constant which can be set
to 1 without loss of generality. This is equivalent to our
normalization condition $\alpha_\vn^{(0)}(\vn)=1$ and
$\alpha_\vn^{(s)}=0$ for $s>1$ in the previous section. We thus obtain
\begin{eqnarray*}
\alpha_\vn = [I + \gamma(\opA - \cE - \gamma \opPperp \opS)^{-1}
\opPperp\opS] \delta_\vn \: ,
\end{eqnarray*}
and the condition in \Ref{AB3} determining the eigenvalue becomes
\begin{eqnarray*}
\cE_\vn = \cE_0(\vn) - \gamma (\opS\alpha_\vn)(\vn) \: .  
\end{eqnarray*}
We expand $\alpha_\vn$ in a geometric series
\begin{eqnarray*} 
\alpha_\vn = \sum_{s=0}^\infty \Bigl( \gamma (\opA-\cE)^{-1}\opPperp
\opS \Bigr)^s \delta_\vn
\end{eqnarray*}
where we introduce the convenient notation
\begin{equation} 
\label{doublebracket}
\frac1{[[ \cE_0(\vm)-\cE]]_{\vn}^{\phantom s}} \define
\left\{ \begin{array}{ll} 
0 & \mbox{ if }\; \vm=\vn \\ 
{[}\cE_0(\vm) -\cE{]}^{-1} & \mbox{ otherwise } 
\end{array} \right. 
\end{equation}
which allows us to write 
\begin{eqnarray*} 
\label{opAcE}
([\opA-\cE]^{-1} \opPperp \alpha)(\vm) = \frac1{[[
\cE_0(\vm)-\cE]]_{\vn}^{\phantom s}} \alpha(\vm) \: . 
\end{eqnarray*}
We thus obtain our first main result.

\begin{theorem}
\label{thm1}
Let $\vn\in\Z^N$ and
\begin{equation}
\label{ansatz1}
\Psi_{\vn}(\vx) = \sum_{\vm\in\Z^N} \alpha_{\vn}(\vm) \hat
F_{\vm}(\vx)
\end{equation}
where 
\begin{eqnarray}
\label{alpha} 
\alpha_{\vn}(\vm) = \delta(\vn,\vm) + \sum_{s=1}^\infty \gamma^s
\sum_{\vnu_1,\ldots,\vnu_s} \prod_{\kappa=1}^s S_{\nu_\kappa}
\frac{\delta(\vm,\vn+\sum_{\kappa =1}^s\vnu_\kappa )}{\prod_{\kappa
=1}^s \Bigl[\!\Big[ \cE_0(\vn + \sum_{\ell=1}^\kappa \vnu_\ell )
-\cE_{\vn} \Bigr]\!\Big]_{\vn}^{\phantom s}}
\end{eqnarray}
with $\hat F_\vn(\vx)$, $\cE_0(\vn)$, $S_\nu$, $\vnu$ defined in
Proposition~\ref{prop1}, and $\cE_{\vn}$ a solution of the following
equation,
\begin{eqnarray} 
\label{cE} 
\cE_{\vn} = \cE_0(\vn) - \sum_{s=2}^\infty \gamma^{s}
\sum_{\vnu_1,\ldots,\vnu_{s} } \prod_{\kappa =1}^{s}
S_{\nu_\kappa} \frac{\delta(\vzero,\sum_{\kappa=1}^{s}
\vnu_\kappa)}{\prod_{\kappa=1}^{s-1} \Bigl[\!\Bigl[ \cE_0(\vn +
\sum_{\ell=1}^\kappa \vnu_\ell) -\cE_{\vn}
\Bigr]\!\Big]_{\vn}^{\phantom s}} \: .
\end{eqnarray}
Then $\Psi_{\vn}(\vx)$ is an eigenfunction of the eCS differential
operator $H$ in \Ref{eCS}--\Ref{V0} with eigenvalue $\cE_{\vn}$:
$H\Psi_{\vn}(\vx) = \cE_{\vn} \Psi_{\vn}(\vx)$.

\end{theorem} 

Our arguments above prove that this result holds true at least in the
sense of formal power series in the nome $q$ of the elliptic
functions, but, as we discuss in the next section, in many cases the
series have a finite radius of convergence.

As before can the sum in \Ref{ansatz1} be restricted to $\vm\in\Z$
satisfying $\sum_{j=1}^N (m_j-n_j)=0$.

Note that \Ref{cE} is an implicit equation determining the eigenvalues
of the eCS model. It is straightforward to obtain from it a solution
as a formal power series in $\gamma$ and thus recover the results in
the previous section. It is important to note that the problem of
resonances has disappeared in \Ref{alpha} and \Ref{cE}, and it only
reemerges if one solves these equations by a particular series
expansion (which is essentially the one discussed in the previous
section). However, it is possible to compute other series solutions
avoiding resonances.

\begin{remark}
\label{rem2} 
It is important to note that \Ref{cE} simplifies in the trigonometric
limit as follows,
\begin{equation} 
\cE_{\vn} = \cE_0(\vn)\; \mbox{ for $q=0$}
\end{equation}
(since for $q=0$, $S_\nu$ is non-zero only for $\nu>0$, but all the
Kronecker deltas in the sums on the r.h.s.\ of \Ref{cE} are zero if
all $\nu_r>0$), and we thus recover the well-known eigenvalues of the
Sutherland model \cite{Su}. Thus in this case, Theorem~\ref{thm1}
provides fully explicit and well-defined eigenfunctions
$\Psi_\vn(\vx)$: one can prove that for all partitions $\vn$,
$\cE_0(\vn+\sum_{\ell=0}^\kappa \vnu_\ell)-\cE_0(\vn) > 0$ for all
$\kappa>0$ and $\nu_\ell >0$ \cite{EL3}, and thus resonances do not
appear in \Ref{alpha} for $q=0$.  As already mentioned, for partitions
$\vn$ and for $q=0$ these eigenfunctions are the same as Sutherland's
\cite{EL3,HL}. In the following we assume $q > 0$.
\end{remark}

We now introduce some useful notation.  We write \Ref{cE} as follows,
\begin{equation}
\label{cE1}  
\tilde\cE_\vn = \Phi_\vn(\tilde\cE_\vn)\; \mbox{ where }\;
\tilde\cE_\vn\define\cE_\vn-\cE_0(\vn)   
\end{equation}
and 
\begin{equation} 
\label{defPhi} 
\Phi_\vn(z) \define - \sum_{s=2}^\infty \gamma^{s}
\sum_{\vnu_1,\ldots,\vnu_{s} } \prod_{\kappa =1}^{s} S_{\nu_\kappa }
\frac{\delta(\vzero,\sum_{\kappa =1}^{s} \vnu_\kappa )}{\prod_{\kappa =1}^{s-1}
\Bigl[\!\Bigl[ \cE_0(\vn + \sum_{\ell=1}^\kappa  \vnu_\ell) -\cE_0(\vn) - z
\Bigr]\!\Big]_{\vn}^{\phantom s}}
\end{equation}
is a complex valued function of one complex variable $z$. Similarly we
write \Ref{alpha} as
\begin{equation}
\label{alpha1} 
\alpha_\vn(\vm) = G_\vn(\tilde\cE_\vn;\vm) 
\end{equation}
where
\begin{equation} 
\label{defG} 
G_\vn(z;\vm) \define \delta(\vn,\vm) + 
\sum_{s=1}^\infty \gamma^s
\sum_{\vnu_1,\ldots,\vnu_s} \prod_{\kappa =1}^s S_{\nu_\kappa }
\frac{\delta(\vm,\vn+\sum_{\kappa =1}^s\vnu_\kappa )}{\prod_{\kappa =1}^s 
  \Bigl[\!\Big[\cE_0(\vn + \sum_{\ell=1}^\kappa  \vnu_\ell ) -\cE_0(\vn) - z
  \Bigr]\!\Big]_{\vn}^{\phantom s}}  
\end{equation}
(for fixed $\vn$ and $\vm$) is also a complex function.

\subsection{Analyticity results} 
\label{sec4.2} 
In this subsection we establish results concerning the analyticity of
the functions $\Phi_\vn(z)$ and $G_\vn(z;\vm)$ introduced at the end
of the previous section.  This is rather technical but needed to prove
convergence of our series solution.

As we show below, for sufficiently small values of $q$, the above
mentioned functions are analytic in certain regions of the complex
$z$-plane. To simplify our discussion we now state a certain
non-degeneracy condition which, when fulfilled, allows us to give a
simple proof of convergence. We emphasis that neither this hypothesis
nor the estimates we deduce from it are optimal, which is why we only
get a proof of convergence in certain special cases.

\begin{hypothesis}
\label{hyp}
The model parameters $N$, $\lambda$ and $\vn\in\Z^N$ are such that
there exist constants $a\in\R$ and $\Delta>|a|$ satisfying
\begin{eqnarray} 
\label{NoR1} 
|\cE_0(\vm )-\cE_0(\vn)-a|\geq \Delta>0 \quad \forall \vm\neq \vn\; \mbox{
  such that }\;  \sum_{j=1}^N(m_j-n_j) =0 \: . 
\end{eqnarray}
\end{hypothesis}

Note that $a$ and $\Delta$ can depend on $\vn$, $N$ and $\lambda$.

It is interesting to note that, for rational values of $\lambda$, it
is trivial to find constants $a\neq 0$ and $\Delta\geq |a|$ such that
the condition in \Ref{NoR1} is fulfilled:

\begin{lemma}
\label{lem0}
Let $\lambda$ be rational and $m$ and $n>0$ co-prime integers such
that $\lambda=n/m$. Then any of the following parameters
\begin{equation} 
\label{rational} 
a = k_1+\lambda k_2 + a_0, \quad k_1, k_2\in\Z \; \mbox{ and }\;
\Delta=|a_0|\leq \frac1{2m}
\end{equation}
obey the condition in \Ref{NoR1}. 
\end{lemma}

\noindent \textit{Proof:} Equation \Ref{Enn} implies that
$\cE_0(\vn)-\cE_0(\vm)$ is always of the form $\nu_1+\lambda \nu_2$
for some integers $\nu_{1,2}$, and thus it is obvious that
$\left|\nu_1+\lambda\nu_2-a\right|\geq \Delta$ for all integers
$\nu_{1,2}$ and the parameters in \Ref{rational}. \QED

This lemma is enough to establish that the functions defined at the
end of Section~\ref{sec4.1} are analytic in non-trivial $z$-regions
for sufficiently small values of $q$
(Proposition~\ref{analyticity}). However, to prove that the
prerequisites for Lagrange's theorem are fulfilled
(Corollary~\ref{corl3}) we need that $\Delta$ is strictly larger than
$|a|$, and it is this which is non-trivial in our hypothesis.  At the
end of this section we prove that Hypothesis~\ref{hyp} is fulfilled
for all $\vn$ if $N=2$ and $\lambda\neq \N$ (Lemma~\ref{lem0a}). For
$N>2$ we only have some partial results, including the groundstate
$\vn=\vzero$ and irrational $\lambda$ (Lemma~\ref{lem0b}).

We now are ready to state our analyticity result:

\begin{prop}
\label{analyticity} 
Let $a\in\R$ and $\Delta>0$ be such that the condition in \Ref{NoR1}
holds true, $b\geq 0$ a parameter which can be chosen arbitrarily, and
$q$ sufficiently small so that
\begin{equation} 
\label{B}
B \define \frac{N(N-1) |\gamma| q^{2/(N+b)}}{(1-q^{2/(N+b)})^3}
\end{equation}
satisfies $B < \Delta$.  Then $\Phi_\vn(z)$ in \Ref{defPhi}
and $G_\vn(z;\vm)$ in \Ref{defG} are analytic functions in the
following region
\begin{equation} 
\label{xi-region} 
|z-a| < \Delta - B
\end{equation}
of the complex $z$-plane.  Moreover, in that region the following
estimates hold true,
\begin{equation} 
\label{Phibound}   
|\Phi_\vn(z)| < \frac{B^2}{\Delta-B-|z-a|}
\end{equation}
and 
\begin{equation} 
\label{Gbound} 
|G_\vn(z;\vm)| < \delta(\vm,\vn) + q^{\sum_j K j(m_j-n_j)}
\frac{B}{\Delta-B-|z-a|} 
\end{equation}
with the constant $K$ in \Ref{KC}.
\end{prop}

\noindent \textit{Proof:} The estimate in \Ref{NoR1} implies
\begin{eqnarray*}
\left|\frac1{\cE_0(\vm)-\cE_0(\vn) - z}\right|\leq
\frac1{|\cE_0(\vm)-\cE_0(\vn)- a|-|z-a|} \leq
\frac1{\Delta-|z-a|}
\end{eqnarray*}
for all $\vm=\vn+\sum_\ell\vnu_\ell\neq \vn$ and $|z-a|<\Delta$, and
thus \Ref{defPhi} yields
\begin{eqnarray*} 
|\Phi_\vn(z)| \leq \sum_{s=2}^\infty
\left(\frac{1}{\Delta-|z-a|} \right)^{s-1} K_{s}(\vzero)
\end{eqnarray*} 
where
\begin{equation} 
\label{KKsN1}
K_s(\vm) \define |\gamma|^s\sum_{\vnu_1,\ldots,\vnu_s} \prod_{\kappa
  =1}^s S_{\nu_\kappa } \delta(\vm ,\mbox{$\sum_{\kappa
    =1}^s\vnu_\kappa$} )>0 
\end{equation}
for all $\vm\in\Z^N$ and $s\in\N$; we used $S_\nu\geq 0$. One can
prove that
\begin{equation}
\label{KsN1}
K_s(\vm) \leq q^{2\sum_j j m_j /(N+b)} B^s 
\end{equation}
with $B$ in \Ref{B} and $b>0$ arbitrary; see
Appendix~\ref{KsNestimate} for details. This implies
\begin{eqnarray*} 
|\Phi_\vn(z)| \leq B \sum_{s=2}^\infty
 \left(\frac{B}{\Delta-|z-a|} \right)^{s-1} 
\end{eqnarray*} 
provided that $|z-a|<\Delta - B$. Summing up the geometric series we
obtain the estimate in \Ref{Phibound}. It is obvious that each term in
the series defining $\Phi_{\vn}(z)$ in \Ref{defPhi} is an analytical
function in the region $|z-a|<\Delta$, and thus our estimates above
prove that $\Phi_\vn(z)$ converges and defines an analytic function in
the region defined in \Ref{xi-region}; see e.g.\ \cite{WW}, Section
5.3.

The proof of analyticity of $G_\vn(z;\vm)$ in \Ref{defG} is similar:
we now can estimate
\begin{eqnarray*} 
|G_\vn(z;\vm)| \leq \delta(\vn,\vm) + \sum_{s=1}^\infty
\left(\frac{1}{\Delta-|z-a|}\right)^s K_s(\vm-\vn)
\end{eqnarray*}
with $K_s(\vm)$ in \Ref{KKsN1}. Using the estimate in \Ref{KsN1} we
obtain
\begin{eqnarray*} 
|G_\vn(z;\vm)| \leq \delta(\vn,\vm) + \sum_{s=1}^\infty q^{2\sum_j
  j(m_j-n_j)/(N+b)} \left(\frac{B}{\Delta-|z-a|}\right)^s
\end{eqnarray*}
which implies the estimate in \Ref{Gbound} provided that
$|z-a|<\Delta-B$. Similarly as above we conclude that $G_\vn(z;\vm)$
is analytic in the region defined in \Ref{xi-region}.\QED

We solve the equation in \Ref{cE1} by using Lagrange's
theorem~\ref{lagthm}.  The following result gives a sufficient
condition that the prerequisites for the stronger version (A) of this
theorem are fulfilled:

\begin{corl}\label{corl3} 
  Let $a\in\R$ and $\Delta>|a|$ be such that the condition in
  \Ref{NoR1} holds true, $b>0$ an arbitrary parameter, and $q$
  sufficiently small so that $B$ in \Ref{B} satisfies $B<
  (\Delta-|a|)/3$. Then for any of the following closed contours in
  the complex $z$-plane,
\begin{equation} 
\label{C} 
\cC:\; |z-a| = \frac12 \left(\Delta-B+|a| - \eps
\sqrt{(\Delta-B-|a|)^2 - 4B^2}\right)\; , \quad -1<\eps<1
\end{equation}
the following condition holds true,
\begin{equation} 
\label{tphiest} 
\sup_{z\in\cC} \frac{|\Phi_\vn(z)-a|}{|z-a|} <1\: ,
\end{equation}
and $\Phi_\vn(z)$ and all functions $G_\vn(z;\vm)$ are analytic inside
and on $\cC$.
\end{corl}

\noindent \textit{Proof:} It follows from the estimate in
\Ref{Phibound} that the condition in \Ref{tphiest} is implied by
\begin{eqnarray*} 
\frac{B^2}{\Delta-B-|z-a|} + |a| < |z-a|\: . 
\end{eqnarray*} 
The latter holds true if and only if
\begin{eqnarray*} 
\left(|z-a| - (\Delta-B+|a|) \right)^2 < \left(\Delta-B-|a|\right)^2 -
4B^2 \: , 
\end{eqnarray*} 
which has non-trivial solutions as in \Ref{C} provided that $2B<
\Delta-B-|a|$.  This proves \Ref{tphiest} under the given
assumptions. The statements concerning analyticity are implied by
Proposition~\ref{analyticity}. \QED

\begin{remark}
\label{rem2_1} 
The best possible bound in \Ref{Phibound} is obviously obtained for
$b=0$, but we need $b>0$ to prove square integrability of the
eigenfunctions.
\end{remark}

To show that Hypothesis~\ref{hyp} is relevant we now discuss a few
special cases where it holds true.

\begin{lemma}\label{lem0a}
  Let $N=2$ and $\lambda$ be non-integer. Then Hypothesis~\ref{hyp}
  holds true with
\begin{equation}
\label{N=2}
a=0,\quad \Delta= \min_{\nu\in\Z'}|2\nu(\nu + n_1-n_2+\lambda)| >0\;
\; \mbox{ if $\lambda\notin\Z$}\; \quad (N=2)\: .
\end{equation} 
\end{lemma} 

\noindent \textit{Proof:} This is a simple consequence of
Equation~\Ref{dE_N=2}. \QED

It is interesting to note that $\Delta$ in \Ref{N=2} can be
arbitrarily large, e.g.\ for $n_1-n_2+\lambda=k + r$, $k\in\N$ and
$0<r<1/2$, $\Delta$ can be as large as $2 r k$. It thus is possible to
find cases where $\Delta$ is large enough for our series to converge
for all values $q<1$.

The reason why we did not find a simple, general proof of convergence
are the resonances discussed at the end of Section~\ref{sec3}, as can
be seen by the following result:

\begin{lemma}
\label{lem0b}
Let $N\geq 2$ and $\lambda$ and $\vn\in\Z^N$ such that the
no-resonance condition in \Ref{NoR} is satisfied. Then
Hypothesis~\ref{hyp} holds true for $a=0$ and some $\Delta>0$.
\end{lemma}

\noindent \textit{Proof:} Since the set of all $\vm$ such that
$|\cE_0(\vm)-\cE_0(\vn)|<1$ (say) is obviously finite,
$$
\Delta \define \inf_{\vm}\left|\cE_0(\vm)-\cE_0(\vn)\right|
$$
can be computed as minimum over a finite set, and thus \Ref{NoR}
implies $\Delta>0$.\QED

Note that the no-resonance condition is satisfied in all cases $\vn$
where there is no $\lambda$-independent resonance and where $\lambda$
is irrational. Another interesting special case where Lemma
\ref{lem0b} applies is the groundstate $\vn=\vzero$ and irrational
$\lambda$.

\begin{remark}
\label{rem3}
The careful reader might wonder why our estimates in
Proposition~\ref{analyticity} are not analytic in $q^2$.  The reason
is that these estimates are not optimal. Indeed, since $\Phi_\vn(z)=0$
for $q=0$ (see Remark~\ref{rem2}), we expect that $\Phi_\vn(z)$ should
vanish like $q^2$ as $q\to 0$. In Remark~\ref{remC1} (Appendix) we
motivate the
\begin{equation}
\label{ConjKsbound} 
\mbox{Conjecture: }\; K_s(\vzero) \leq \tilde{B}^{s} 
q^{2\lceil\frac{s}{N}\rceil}\; \mbox{ with }\;
\tilde{B}\define \frac{N(N-1) |\gamma|}{(1-q^{2/N})^3} \: . 
\end{equation} 
This would imply the following improved estimate
\begin{eqnarray*}
\label{ConPhibound} 
\mbox{Conjecture: }\; 
|\Phi_\vn(z)|\leq \tilde{B} q^2 
\left( \frac{(\Delta-|z-a|)}{[(\Delta-|z-a|)^N-
    \tilde{B}^N q^2]}\sum_{n=1}^{N-1} \tilde{B}^{N-n}(\Delta-|z-a|)^n  
  -1 \right) 
\end{eqnarray*} 
analytic in $q^2$ and consistent with $\Phi_\vn(z)=\cO(q^2)$. We
expect there exists a similar improved estimate for $G_\vn(z;\vm)$.
\end{remark} 

\subsection{Explicit solution} 
\label{sec4.3} 
From the implicit equation for $\cE_\vn$ in Theorem~\ref{thm1} it is
straightforward to get an explicit expression by expanding in $\gamma$
and thus extend the series solution in Section~\ref{sec3} to all
orders. It is important to note that this amounts to an expansion
around $\cE=\cE_0(\vn)$. As will be shown below, it is possible to
expand around any point $\cE=\cE_0(\vn)+a$, $a$ arbitrary, and thus
the `resonance denominators' $\cE_0(\vm)-\cE_0(\vn)$ can be moved to
$\cE_0(\vm)-\cE_0(\vn)-a$. In this way the resonance problem can be
circumvented, and we obtain the following fully explicit result.

\begin{theorem}
\label{thm2}
(A) Let $\vn\in\Z^N$ and $a\in\R$ such that the condition in
\Ref{NoR1} holds true for some $\Delta>0$. Then the eigenvalue
equation $H\Psi(\vx)=\cE\Psi(\vx)$ of the eCS differential operator in
\Ref{eCS}--\Ref{V0} has a solution $\cE=\cE_\vn$ as follows,
\begin{eqnarray} 
  \label{xi00} 
\cE_\vn = \cE_0(\vn) + a + \sum_{m=1}^\infty
\sum_{\ell_0,\ell_1,\ldots,\ell_{m-1}=0}^\infty
\delta(\mbox{$\sum_{r=0}^{m-1}$} \ell_r, m) \nonu \times
\delta(\mbox{$\sum_{r=1}^{m-1}$} r \ell_r, m-1) (m-1)!
\prod_{r=0}^{m-1} \frac{[\Phi^{(r)}_\vn(a)]^{\ell_r}}{\ell_r!}
\end{eqnarray}
with 
\begin{eqnarray} 
\label{Phir}
\Phi^{(r)}_\vn(a) = -\sum_{s=2}^\infty \gamma^{s}
\sum_{\vnu_1,\ldots,\vnu_{s} } \prod_{\kappa=1}^{s} S_{\nu_\kappa}
\sum_{k_1,k_2,\ldots,k_{s-1}=0}^\infty
\delta(\mbox{$\sum_{\kappa=1}^{s-1}$} k_\kappa,r) \nonu \times
\frac{\delta(\vzero,\sum_{\kappa=1}^{s}\vnu_\kappa)}{\prod_{\kappa=1}^{s-1}
\Bigl[\!\Bigl[ \cE_0(\vn + \sum_{\ell=1}^\kappa \vnu_\ell) -\cE_0(\vn)
-a \Bigr]\!\Big]_{\vn}^{1+k_\kappa}} - \delta_{r,0} a
\end{eqnarray}
and the notation defined in \Ref{Ejk}, \Ref{vnu} and
\Ref{doublebracket}. Moreover, the coefficients $\alpha_\vn(\vm)$
giving the corresponding eigenfunction $\Psi(\vx)=\Psi_{\vn}(\vx)$
according to Theorem~\ref{thm1} are
\begin{eqnarray} 
\label{a00} 
\alpha_\vn(\vm) = G_\vn^{(0)}(a;\vm) + \sum_{m=1}^\infty
\sum_{\ell=0}^{m-1} \sum_{\ell_0,\ell_1,\ldots,\ell_{\ell}=0}^\infty
\delta(\mbox{$\sum_{r=0}^{\ell}$} \ell_r, m)
\delta(\mbox{$\sum_{r=1}^{\ell}$} r \ell_r, \ell)\nonu \times
(m-1)!(m-\ell) G_\vn^{(m-\ell)}(a;\vm)
\prod_{r=0}^{\ell}\frac{[\Phi_\vn^{(r)}(a)]^{\ell_r}}{\ell_r!}
\end{eqnarray}
with $\Phi^{(r)}_\vn(a)$ as above and
\begin{eqnarray} 
\label{Gr} 
G_\vn^{(r)}(a;\vm) = \delta_{r,0}\delta(\vm,\vn) + \sum_{s=1}^\infty
\gamma^s \sum_{\vnu_1,\ldots,\vnu_s} \prod_{r=1}^s S_{\nu_r}
\sum_{k_1,k_2,\ldots,k_s=0}^\infty \delta(\mbox{$\sum_{\kappa=1}^{s}$}
k_\kappa,r) \nonu \times
\frac{\delta(\vm,\vn+\sum_{\kappa=1}^s\vnu_\kappa)}{\prod_{\kappa=1}^s
\Bigl[\!\Big[ \cE_0(\vn + \sum_{\ell=1}^\kappa \vnu_\ell )
-\cE_0(\vx)-a \Bigr]\!\Big]_{\vn}^{1+k_\kappa}} .
\end{eqnarray}
The results above hold true for arbitrary parameter values, and
$\cE_\vn$ and $\alpha_\vn(\vm)$ above are independent of the parameter
$a$ in the sense of formal power series.

\noindent (B) If $\Delta>|a|$ and if $q$ is small enough that $B$ in
\Ref{B} satisfies the condition $B< (\Delta-|a|)/3$, then all series
above converge absolutely, the following estimates holds true,
\begin{equation}
\label{Enbound} 
|\cE_\vn-\cE_0(\vn)-a| \leq 
\frac12 \left(\Delta-B+|a|
- \sqrt{(\Delta-B-|a|)^2 - 4B^2}\right)
\end{equation}
and
\begin{equation} 
\label{anmbound} 
|\alpha_\vn(\vm)| \leq \delta(\vm,\vn) + q^{\sum_j j K (m_j-n_j) }
\frac{2B}{\Delta-B-|a| +  \sqrt{(\Delta-B-|a|)^2 - 4B^2} }  \: , 
\end{equation}
with $K$ in \Ref{KC}, 
and the eigenfunction $\Psi_{\vn}(\vx)$ determined by the formulas
above is square integrable.
\end{theorem} 

\noindent \textit{Proof: } We observe that \Ref{cE1} is of the form
$z=\eta\varphi(z)$ if we identify 
\begin{equation} 
\label{defz}
z = \tilde\cE_\vn\: , \quad \eta \varphi(z) = \Phi_\vn(z)\: . 
\end{equation} 
We thus can use Lagrange's theorem \ref{lagthm} to solve
\Ref{cE1}. Note that $\eta$ can be regarded as a book keeping
parameter which can be set to any value. We set $\eta=1$ in the
following.

We first prove the result under the assumptions stated in (B).
According to Corollary~\ref{corl3} the prerequisites for Lagrange's
theorem are then fulfilled for any of the contours defined in
\Ref{C}. Thus the equation $z=\Phi_\vn(z)$ has a single, simple
solution $z=\xi$ within any such contour. Taking the infimum over the
allowed $\varepsilon$-values we obtain the estimate in
\Ref{Enbound}. Moreover, using \Ref{Lagrange1} for $g(z)=z$ we obtain
the following explicit formula for the eigenvalues by an absolutely
convergent series,
\begin{eqnarray*} 
\cE_\vn=\cE_0(\vn)+\xi\; \mbox{ with }\; \xi = a+ \sum_{m=1}^\infty
\frac{1}{m!}  \frac{\dd^{m-1}}{\dd a^{m-1}} \varphi(a)^m\: . 
\end{eqnarray*}
We now can write the equation in \Ref{alpha1} as
\begin{equation} 
\label{defa} 
\alpha_\vn(\vm) = g(\xi)\: ,\quad g(z)= G_\vn(z;\vm)
\end{equation}
where $g(z)$ fulfills the prerequisites of Lagrange's
theorem~\ref{lagthm} according to Corollary~\ref{corl3}. Thus
\Ref{Lagrange1} gives an explicit formula for the $\alpha_\vn(\vm)$ by
an absolutely convergent series. Moreover, inserting \Ref{Enbound} in
the estimate in \Ref{Gbound} we obtain \Ref{anmbound}.

To make this series more explicit we insert \Ref{phi-series} and use
the following multinomial series,
\begin{eqnarray} 
\label{multinomial}
\Bigl( \sum_{r=0}^\infty \varphi_r (z-a)^r \Bigr)^m =
\sum_{\ell_0,\ell_1,\ldots=0}^\infty\delta(\mbox{$\sum_{r=0}^\infty$}\ell_r,m)
\frac{m!}{\prod_{r=0}^\infty \ell_r! }\prod_{r=0}^\infty
[\varphi_r]^{\ell_r} (z-a)^{\sum_{r=0}^\infty r \ell_r}= \nonu
\sum_{\ell=0}^\infty (z-a)^\ell \sum_{\ell_0,\ldots,\ell_\ell
  =0}^\infty\delta(\mbox{$\sum_{r=0}^\ell$}\ell_r,m)
\delta(\mbox{$\sum_{r=1}^\ell$}r \ell_r,\ell)
\frac{m!}{\prod_{r=0}^\ell \ell_r!} \prod_{r=0}^\ell
[\varphi_r]^{\ell_r}\: , 
\end{eqnarray}
and similarly for $g$. This yields
\begin{eqnarray} 
\label{xi22} 
\xi = a+\sum_{m=1}^\infty
\sum_{\ell_0,\ell_1,\ldots,\ell_{m-1}=0}^\infty
\delta(\mbox{$\sum_{r=0}^{m-1}$} \ell_r, m)
\delta(\mbox{$\sum_{r=1}^{m-1}$} r \ell_r, m-1)(m-1)!\prod_{r=0}^{m-1}
\frac{[\varphi_r]^{\ell_r}}{\ell_r!}
\end{eqnarray}
and
\begin{eqnarray} 
\label{alpha22} 
g(\xi) = g(a) + \sum_{m=1}^\infty \sum_{\ell=0}^{m-1}
\sum_{\ell_0,\ell_1,\ldots,\ell_{\ell}=0}^\infty
\delta(\mbox{$\sum_{r=0}^{\ell}$} \ell_r, m) \nonu \times
\delta(\mbox{$\sum_{r=1}^{\ell}$} r \ell_r, \ell)
(m-1)!(m-\ell)g_{m-\ell}
\prod_{r=0}^{\ell}\frac{[\varphi_r]^{\ell_r}}{\ell_r!}\: .
\end{eqnarray}
We observe that
\begin{equation} 
\varphi_r = \frac1{r!} \frac{\dd^r}{\dd a^r} \left( 
\Phi_\vn(a) -a\right) \define \Phi_\vn^{(r)}(a)\: .
\end{equation}
Recalling \Ref{defz} and inserting \Ref{defPhi} we obtain \Ref{xi00}
and \Ref{Phir}; we used
\begin{equation}
\label{Leibniz}
\frac1{r!}\frac{\dd^r}{\dd a^r} \frac1{\prod_{\kappa=1}^s(x_\kappa-a)}
= \sum_{k_1,\ldots,k_s=0}^\infty \delta(\mbox{$\sum_{\kappa=1}^s
k$}_\kappa,r) \frac1{\prod_{\kappa=1}^s(x_\kappa-a)^{1+k_\kappa} }
\end{equation}
for complex parameters $x_\kappa$, which follows from the Leibniz
rule.  Similarly, recalling
\begin{equation}
g_r= \frac1{r!} \frac{\dd^r}{\dd
a^r} G_\vn(a;\vm) \define G_\vn^{(r)}(a;\vm) 
\end{equation}
and \Ref{defa} and using \Ref{defG} we obtain \Ref{a00} and \Ref{Gr}.

The estimates in \Ref{anmbound} and Lemma~\ref{lemPn} imply
\begin{eqnarray*}
|\Psi_\vn(\vx)|\leq \sum_{\vm}|\alpha_\vn(\vm)||f_\vm(\vz)| < 
\sum_\vm  \Bigl( \delta(\vm,\vn) + q^{\sum_j j K (m_j-n_j) }\tilde C\Bigr) 
\times \nonu C \, q^{\sum_j (\tilde{K} |m_j| - K j m_j) } 
= 
  C \, q^{-\sum_j K j n_j }
  \Biggl( q^{\sum_j\tilde{K} |n_j|} + \tilde C 
    \Bigl(\frac{1+q^{\tilde{K}}}{1-q^{\tilde{K}}} \Bigr)^N \Biggr)
\end{eqnarray*}
for all $b>0$, with $\tilde C= 2B/[\Delta-B-|a| +
\sqrt{(\Delta-B-|a|)^2 - 4B^2} ]$ and the constants in \Ref{KC} and
\Ref{B}. This proves that $\Psi_\vn(\vx)$ is uniformly bounded, and
thus square integrable, on its domain $[-\pi,\pi]^N$. This completes
the proof of (B).

If we only know that $a$ is such that \Ref{NoR1} holds true for some
$\Delta>0$ we cannot conclude anything about convergence. However, the
results stated in (A) still hold true in the sense of formal power
series; see Theorem~\ref{lagthm} and Remark~\ref{remLagrange}. \QED

\begin{remark}
\label{rem5}
Since $\Phi_\vn(z)$ is proportional to $q^2$ (see Remark~\ref{rem3})
one can choose the parameter $\eta$ in \Ref{defz} to be $q^2$, and
this shows that the $m$-th term in \Ref{xi00} and \Ref{a00} is
$\cO(q^{2m})$. From this we expect that these series have a finite
radius of convergence in $q^2$ in general. However, since the function
$\varphi(z)$ also depends on $q^2$, this is not easy to prove.
\end{remark}

\begin{remark}
\label{rem5a}
The result in Theorem~\ref{thm2} explains why the power series of
$\cE_\vn$ in $q^2$ is very complicated: due to \Ref{cycle} the term
$\propto\gamma^s$ on the r.h.s\ in \Ref{defPhi} is
$\cO(q^{2\lceil\frac{s}{N}\rceil})$; see Remark~\ref{remC1} in the
Appendix. Thus all such terms $\propto \gamma^s$ with $s=2,\ldots,\ell
N$ contribute to the power series coefficient $\propto q^{2\ell}$ of
$\cE_\vn$, and the complexity of this $q^{2\ell}$-term therefore
increases dramatically not only with $\ell$ but also with $N$.
\end{remark}

\begin{remark} 
\label{rem6} 
Our proof of convergence relies on estimates in Section~\ref{sec4.2}
which are crude.  In particular, we estimate the energy denominators
$[\cE_0(\vm)-\cE_0(\vn)-z]$ in \Ref{defPhi} and \Ref{defG} by their
smallest possible value which, if resonances exist, are assumed for
resonances. However, resonances are rare, and it is easy to see that
there always exists a constant $\Delta_0>0$ such that
$|\cE_0(\vm)-\cE_0(\vn)|>\Delta_0$ for all $\vm$ different from a
resonance.  It thus seemed to us that it should be easy to improve our
estimates in Proposition~\ref{analyticity} enough to prove convergence
of our series for arbitrary particle numbers $N$.  However, despite of
much effort (delaying the publication of this paper for two more
years) we have not been able to do this up to now. Anyway, we have
several independent reasons to believe in the existence of a finite
radius of convergence for arbitrary parameters: (i) the results in
\cite{KT}, (ii) the existence of variants of our solution method
described in Section~\ref{sec4.4} and Remark~2 in Section~\ref{sec5}
which both avoid resonances, (iii) recent numerical results \cite{BL}.
\end{remark}

\subsection{A method to avoid resonances}
\label{sec4.4} 
We now describe a generalization of our solution described in
Sections~\ref{sec4.1}--\ref{sec4.3} which allows to avoid
resonances. The idea is to use a variant of degenerate perturbation
theory.

We fix $\vn\in \Z^N$ and consider the set of all corresponding
resonances, i.e.\ all $\vm\in\Z^N$ different from $\vn$ such that
$\sum_j(m_j-n_j)=0$ and $\cE_0(\vm)=\cE_0(\vn)$. It is easy to see
that there can be only a finite number of distinct resonances which we
denote as $\vn_2$, $\vn_3$, $\ldots$, $\vn_{R+1}$ with $R$ the number
of resonances. We also set $\vn_1\define \vn$. For example, for $N=2$
and integer $\lambda$ we get $R=1$, $\vn_1=(n_1,n_2)$ and
$\vn_2=(n_2-\lambda,n_1+\lambda)$; see \Ref{dE_N=2}.

We now apply Lemma~\ref{SolutionLemma} to \Ref{eq3} using the
following projection, $(\opP_\vn\alpha)(\vm) \define \alpha(\vm)$ for
$\vm=\vn_J$, $J=1,2,\ldots,R+1$, and $0$ otherwise; $\opA$ and $\opB$
are as before.  The general solution of \Ref{alpha0} is now $\alpha_0
= \sum_{J=1}^{R+1} c_J \delta_{\vn_J}$ with constants $c_J$ to be
determined. This implies the following generalization of
\Ref{ansatz1},
\begin{equation}
\label{ansatz1_1}
\Psi_{\vn}(\vx) = \sum_{J=1}^n \sum_{\vm\in\Z^N} c_J \alpha_{\vn_J}(\vm) \hat
F_{\vm}(\vx)
\end{equation}
with $\alpha_{\vn_J}(\vm)$ obtained by setting $\vn=\vn_J$ in
\Ref{alpha} and
\begin{equation} 
\label{doublebracket_1}
\frac1{[[ \cE_0(\vm)-\cE]]_{\vn}^{\phantom s}} \define
\left\{ \begin{array}{ll} 
    0 & \mbox{ if }\; \vm=\vn_J\; \mbox{ for $J=1,2,\ldots,R+1$}  \\ 
    {[}\cE_0(\vm) -\cE{]}^{-1} & \mbox{ otherwise }\; .  
\end{array} \right. 
\end{equation}
The equation determining the eigenvalues
$\cE_\vn=\cE_0(\vn)+\tilde\cE_\vn$ and the coefficients $c_J$ is
obtained from \Ref{AB2} which can be written as follows,
\begin{equation}
\label{cE_1} 
\tilde\cE_\vn c_J = \sum_{K=1}^{R+1} \Phi_{\vn_J,\vn_K}(\tilde \cE_\vn) c_K  
\end{equation} 
with 
\begin{equation}
\Phi_{\vn_J,\vn_K}(z) =  - \sum_{s=1}^\infty \gamma^{s}
\sum_{\vnu_1,\ldots,\vnu_{s} } \prod_{\kappa =1}^{s}
S_{\nu_\kappa} \frac{\delta(\vn_J, \vn_K + \sum_{\kappa=1}^{s}
\vnu_\kappa)}{\prod_{\kappa=1}^{s-1} \Bigl[\!\Bigl[ \cE_0(\vn_K +
\sum_{\ell=1}^\kappa \vnu_\ell) -\cE_0(\vn)-z 
\Bigr]\!\Big]_{\vn}^{\phantom s}} \: .
\end{equation} 
It should be possible to solve \Ref{cE_1} explicitly using a matrix
version of Lagrange's reversion theorem. One now can expand about
$z=0$, and we expect that our estimates in Section~\ref{sec4.2} can be
made sharp enough to prove absolute convergence in a finite
$q$-interval without restrictions on parameters. It would be
interesting to explore this generalized solution algorithm in more
detail, but this is beyond the scope of the present paper.

\newsection{Final remarks}
\label{sec5}
\noindent \textbf{1.} The functions $\cJ_\vn(\vz)$ defined in \Ref{cJcP}
can be expanded as follows,
\begin{equation}
\cJ_\vn(\vz) = \sum_{\ell=0}^\infty \cJ_{\ell,\vn}(\vz)q^{2\ell}  
\end{equation}
where $\cJ_{0,\vn}(\vz)$ are (essentially) the Jack polynomials
\cite{McD,St}, as discussed in the introduction. It is interesting to
note that for partitions $\vn$, all functions
\begin{equation}
(z_1z_2\cdots z_N)^\ell \cJ_{\vn,\ell}(\vz),\quad \ell=0,1,2,\ldots 
\end{equation}
are symmetric polynomials \cite{EL1}. It would be interesting to
investigate if these polynomials have a combinatorial significance
also for $\ell\geq 1$.

\noindent \textbf{2.} Proposition \ref{prop1} above suggest that the
function in \Ref{F} has an expansion in eigenfunctions $\Psi_\vn(\vx)$
of the eCS Hamiltonian as follows,
\begin{equation} 
F(\vx;\vy) = \sum_{\vn} \kappa_{\vn}\Psi_\vn(\vx)
\overline{\Psi_\vn(\vy)} 
\end{equation}
for some constants $\kappa_{\vn}$ and the bar indicating complex
conjugation, and our algorithm provides a means to extract from this
the eigenfunctions.

It is well-known that the eCS model has a family of $N$ independent
(formally) self-adjoint differential operators of the form $H_k =
(-\ii)^N\sum_{j=1}^N \frac{\partial^k}{\partial x_j^k}+$(lower order
terms) for $k=1,2,\ldots N$ which mutually commute, $[H_k,H_\ell] = 0$
for all $k,\ell=1,2,\ldots,N$, and including the total momentum
operator $P=H_1$ and the eCS Hamiltonian $H=H_2$ in \Ref{eCS}
\cite{OP}. This suggests that the remarkable identity should be
generalizable to all these differential operators,
\begin{equation} 
\label{Hk} 
\mbox{Conjecture: } \quad [ H_k(\vx) - H_k(\vy) ] F(\vx;\vy) = 0
\; \mbox{ for all $k$}  
\end{equation}
(for $k=1$ the proof is trivial). Using this one could extend our
results to the operator $\sum_{k=1}^N b_k H_k$ for arbitrary real
coefficients $b_k$. The resulting formulas would be similar to ours
but with $\cE_0(\vn)$ in \Ref{Enn} replaced by the $\sum_{j,k=1}^N
b_k\left( n_j + \frac{\lambda}{2}(N+1-2j)\right)^k$. The freedom to
choose the parameters $b_j$ arbitrarily should allow to avoid the
resonance problem completely, similarly as in \cite{KT}. It would be
interesting to find a direct proof of the identities in \Ref{Hk} and
to derive and explore the explicit solution obtained with the operator
$\sum_{k=1}^N b_k H_k$.

\noindent \textbf{3.} To set our results in perspective we note that a
\textit{formal} series representation of eigenfunctions and
eigenvalues of the kind derived in this paper can be given for any
quantum mechanical model. Indeed, assume that we want to diagonalize
some self-adjoint Hilbert space operator $H$ using some countable
generating set $f_n$ of the pertinent Hilbert space such that
\begin{equation}
H f_n = E_n f_n + \sum_{m} V_{nm}f_m\; \quad (V_{nn}=0) 
\end{equation} 
for some constants $E_n$ and $V_{nm}$. For example, if $H$ is a
Hamiltonian and $f_n$ a complete orthonormal basis then $\langle f_n,H
f_m\rangle=E_n$ for $m=n$ and $V_{nm}$ otherwise, but what we describe
here is equally true for any generating sets which are not orthonormal
and/or are overcomplete. It is easy to see that our arguments in
Section~\ref{sec4} can be generalized to the present general case and
imply that $H$ has formal eigenfunctions
\begin{equation}
\label{BW1} 
\psi_n = \sum_m \alpha_n(m)f_m
\end{equation}
with corresponding formal eigenvalues $\cE_n=E_n+\tilde\cE_n$ where
$\tilde\cE_n$ is a solution of the equation
$\tilde\cE_n=\Phi_n(\tilde\cE_n)$ with
\begin{equation}
\label{BW2} 
\Phi_n(z) = \sum_{s=1}^\infty (-1)^s \sum_{k_1,k_2,\ldots,k_s} 
\frac{V_{nk_1}V_{k_1k_2} \cdots V_{k_s n}}{\prod_{r=1}^s[[E_{k_r}-E_n-z]]_n} 
\end{equation}
where $1/[[E_k-z]]_n=1/[E_k-z]$ for $k\neq n$ and $0$ for
$k=n$. Moreover, the coefficients for these eigenfunctions are given by
\begin{equation}
  \alpha_n(m)=\delta_{nm} + \sum_{s=1}^\infty (-1)^s \sum_{k_0,k_1,\ldots,k_s} 
  \frac{\delta_{n,k_0}\left( 
      \prod_{r=1}^sV_{k_{r-1,k_r}}\right)\delta_{k_s,m}}{\prod_{r=1}^s
    [[E_{k_r}-E_n-\tilde\cE_n]]_n} 
\end{equation} 
(this result is known as Brillouin-Wigner perturbation theory;
see e.g.\ \cite{Lowdin} and references therein).  Using the formula in
\Ref{Lagrange1} one can obtain from this fully explicit formulas for
the eigenvalues and eigenfunctions as formal series to all
orders; see the formulas given in the proof of Theorem~\ref{thm1}. 
This shows that  it is not the existence of an explicit series
solution which makes the eCS model special, but it is rather the
details of this solution. To be more specific, the eCS model is
special since there exists a basis $f_n$ such that $E_n$ and $V_{nm}$
are given by simple explicit formulas, that $V_{nm}$ only depends on
$n-m$, and that the matrix $V_{nm}$ is close to triangular which
implies that the series solution converges (the latter property we
only proved here in special cases, but we believe that it is true in
general).

It is worth mentioning that the formulas above shed some interesting
light on quantum mechanical perturbation theory in general. In
particular, they highlight that, for a given Hamiltonian $H$, it is
the choice of the basis $f_n$ which determines the usefulness of
perturbation theory. Moreover, the formulas above hold true for any
linear operator $H$ and can be easily generalized to degenerate cases
as described in Section~\ref{sec4.4}.

\noindent \textbf{4.} Our algorithm was based on the remarkable identity
in \Ref{rem1} which we first obtained using quantum field theory
techniques \cite{EL1,EL2}. While we latter found an elementary proof
of this result (presented in Appendix~\ref{appA3}), we feel that the
quantum field theory proof is more illuminating since it not only
shows that this result it true but also why. We also note that there
exist interesting generalizations of this identity which we found
using these quantum field theory results \cite{EL6}.

\noindent \textbf{5.} To judge the usefulness of our results it is
important study our series solution numerically. Numerical results for
$N=2$ were recently obtained by J.C.\ Barba and show that the
numerical convergence is much better than the estimates in
Section~\ref{sec4.2} suggest \cite{BL}. To be more specific, it seems
that resonances are no problem in practice, and low order truncations
of our series approximate the exact solution well not only for small
$q$-values but up to values close to $q=1$ and for a wide range of
coupling parameters.

\noindent \textbf{6.} For $\lambda=1$, $q=0$, and partitions $\vn$ the
functions $f_\vn(\vz)$ in \Ref{fn} are identical with the Schur
polynomials \cite{HL}. Our results therefore include simple explicit
formulas for an elliptic deformation of the Schur polynomials which
seem worth to be studied in more detail.

\noindent \textbf{7.} We believe that the solution method presented in
this paper can be extended to other examples, including the anharmonic
oscillator and the Heun equation.\footnote{This was suggested to us by
  Vadim Kuznetsov.} We hope to come back to this in future work.

\noindent \textbf{Acknowledgments.}  This work owes much to Vadim
Kuznetsov and is therefore dedicated to his memory. I am also grateful
to J.C.\ Barba, M.\ Halln\"as, G.\ Lindblad, J.\ Mickelsson, A.\
Polychronakos, S.G.\ Rajeev, S. Ruijsenaars, and E.\ Sklyanin for
helpful comments and to J.C.\ Barba and M.\ Halln\"as for reading the
manuscript.  I thank F.\ Calogero, B.\ Kupershmidt, N.\ Nekrasov, Y.\
Suris, S.\ Rauch, E.\ Sklyanin, and P.\ Winternitz for their interest
and encouragement.  We also acknowledge financial support by the
Swedish Science Research Council (VR), the G\"oran Gustafsson
Foundation, and the European Union through the FP6 Marie Curie RTN
\textit{ENIGMA} (Contract number MRTN-CT-2004-5652).

%%%%%%%%%%%%%%%%%%%%%%%%%%%%%%%%%%%%%%%%%%%%%%%%%%%%%%%%%%%%%%%%%%%%%%%%

\appendix 
\appsection{Identities of elliptic functions} 
\label{appA}
In this Appendix we give an elementary proof of the Fact stated in
\Ref{F}--\Ref{rem} and on which our algorithm is based. For the
convenience of the reader we also include the proofs of some
properties of elliptic functions which we need.

\subsection{Relation of  $V$ and $\wp$} 
\label{appA1}
Here we prove \Ref{V0} and \Ref{V}.

From the definition in \Ref{V0} it is obvious that $V(z)$, $z\in\C$,
is doubly periodic with periods $2\omega_1=2\pi$ and
$2\omega_2=\ii\beta$, it has a single pole of order 2 in each
period-parallelogram, $V(z)-z^{-2}$ is analytic in some neighborhood
of $z=0$ and equal to $c_0$ in \Ref{c0}; we used $q= \exp(\pi\ii\,
\omega_2/\omega_1)=\exp(-\beta/2) $.  These facts imply the second
identity in \Ref{V0} (see e.g.\ \cite{EMOT}, Section~13.12).

To prove \Ref{V} we note that $\tet(2z)$ equals, up to a constant, the
Jacobi Theta function $\vartheta_1(z)$: 
\begin{eqnarray} 
\label{Theta1} 
\theta(z) = \frac{1}{ 2 q^{1/4} \prod_{n=1}^\infty (1-q^{2n})}\: 
\vartheta_1(z/2) 
\end{eqnarray}
(see e.g.\ page 470 in \cite{WW}). From the relation between
$\vartheta_1$ and the Weierstrass elliptic functions $\sigma$, $\zeta$
and $\wp$ we therefore conclude (see e.g.\ page 473 in \cite{WW})
\begin{equation} 
\label{zeta}
\zeta(z) = \frac{d}{dz}
\log\tet(z) +\frac{\eta_1 z}{\omega_1} 
\end{equation} 
and 
\begin{equation} 
\label{wp}
\wp(z) = - \frac{d}{dz}\zeta(z)
= - \frac{d^2}{dz^2} \log\tet(z) - \frac{\eta_1}{\omega_1} 
\end{equation} 
where $\eta_1/\omega_1$ is a constant.  To determine the latter
constant we use the definition in \Ref{tet} and compute
\begin{eqnarray*} 
\log\tet(z) = \log[(z/2) -(z/2)^3/6] + \sum_{n=1}^\infty
\log[(1-q^{2n})^2 + q^{2n}\, z^2 ] +\cO(z^4) =\nonu const.\ + \log(z)
- \left( \frac{1}{24} - \sum_{n=1}^\infty \frac{q^{2n}}{(1-q^{2n})^2}
\right) \, z^2+\cO(z^4) \: .
\end{eqnarray*}
Recalling that $\wp(z)-z^{-2}$ vanishes for $z\to 0$ one concludes
from this and \Ref{wp} that $\eta_1/\omega_1 = c_0$, recalling
\Ref{c0}.  This together with \Ref{V0} and \Ref{wp} proves
\Ref{V}. \QED

\subsection{Proof of the identities in \Ref{rel}--\Ref{defrel}}
\label{appA2}
We start with the following well-known identity for the Weierstrass
elliptic functions $\zeta$ and $\wp$,
\begin{equation}
\label{1}
[\zeta(x) +\zeta(y) + \zeta(z)]^2 = \wp(x) + \wp(y) + \wp(z) \quad
\; \mbox{ if $x+y+z=0$} 
\end{equation} 
(this identity is given as an exercise on page 446 in \cite{WW}). From
\Ref{zeta} we conclude that $\phi(x)=\tet'(x)/\tet(x)$ equals
$\zeta(x)$ up to a term linear in $x$. Thus the identity in \Ref{1}
remains true if we replace $\zeta$ by $\phi$. This together with the
trivial identity
\begin{eqnarray*}
\phi(x)\phi(y) +\phi(x)\phi(z)+ \phi(y)\phi(z) = \nonu = 
\half[\phi(x)+\phi(y)+\phi(z)]^2 - 
\half[\phi(x)^2+\phi(y)^2+\phi(z)^2] 
\end{eqnarray*}
implies \Ref{rel}. \QED

\subsection{Proof of Lemma~\ref{fact1}}
\label{appA3}
Let $F= F(\vx;\vy)$ as defined in \Ref{F}. We compute
\begin{eqnarray*}
\frac{\partial}{\partial x_j} F = \biggl[ \sum_{k\neq
    j}\lambda\phi(x_j-x_k) -\sum_{k}\lambda\phi(x_j-y_k) \biggr] F
\end{eqnarray*}
with $\phi(x)= \tet'(x)/\tet(x)$, and thus
\begin{eqnarray*}
\frac{\partial^2}{\partial x_j^2} F = \biggl[ 
 \sum_{k\neq j}\lambda\phi'(x_j-x_k)-
\sum_{k}\lambda\phi'(x_j-y_k) + 
\nonu
 \sum_{k,\ell\neq j}\lambda^2\phi(x_j-x_k)\phi(x_j-x_\ell)
+ \sum_{k,\ell}\lambda^2\phi(x_j-y_k)\phi(x_j-y_\ell)
\nonu
- 2 \sum_{k\neq j,\ell}\lambda^2\phi(x_j-x_k)\phi(x_j-y_\ell)
\biggr] F \: .
\end{eqnarray*}
With that we compute straightforwardly 
\begin{eqnarray*}
W \, \define \, \frac{1}{F}\sum_{j=1}^N 
\biggl( \frac{\partial^2}{\partial x_j^2} -  \frac{\partial^2}{\partial y_j^2}
\biggr) F 
\end{eqnarray*}
which we write as a sum of four terms, $W=W_1+W_2+W_3+W_4$, 
with
\begin{eqnarray*}
W_1 \define \sum_j 
\sum_{k\neq j}\biggl[\lambda\phi'(x_j-x_k)
+ \lambda^2 \phi(x_j-x_k)^2 \biggr] -[x\leftrightarrow y]
\end{eqnarray*}
(`$[x\leftrightarrow y]$' means the same terms but with the arguments
$x_j$ and $y_j$ interchanged),
\begin{eqnarray*}
W_2 \define \sum_{k,j}\biggl[-\lambda\phi'(x_j-y_k)
+ \lambda^2 \phi(x_j-y_k)^2 \biggr] -[x\leftrightarrow y]\, ,
\end{eqnarray*}
\begin{eqnarray*}
W_3 \define  \sum_j \sum_{k\neq j}\sum_{\ell\neq j,k}\biggl[
\lambda^2 \phi(x_j-x_k)\phi(x_j-x_\ell)\biggr] -[x\leftrightarrow y]\, , 
\end{eqnarray*}
and 
\begin{eqnarray*}
W_4 \define  \sum_{j,k} \sum_{\ell \neq k}\biggl[
\lambda^2 \phi(x_j-y_k)\phi(x_j-y_\ell)
- 2\lambda^2 \phi(x_k-x_\ell)\phi(x_k-y_j)\biggr] -[x\leftrightarrow y]
= \nonu
 \sum_{j,k} \sum_{\ell \neq k}\biggl[
\lambda^2  \phi(x_j-y_k)\phi(x_j-y_\ell)
+ 2\lambda^2 \phi(y_k-y_\ell)\phi(y_k-x_j)\biggr] 
-[x\leftrightarrow y] \: . 
\end{eqnarray*}
We first observe that the first two terms in $W_2$ are invariant under
$x\leftrightarrow y$ [note that $\phi'(-x)=\phi'(x)$], and therefore
\begin{eqnarray*}
W_2=0\: . 
\end{eqnarray*}
We then write $W_3$ as follows [using $\phi(-x)=-\phi(x)$]
\begin{eqnarray*}
W_3 =  \sum_{j<k<\ell}(-\lambda^2) \biggl[
\phi(x_k-x_j)\phi(x_j-x_\ell)+ 
\phi(x_\ell-x_k)\phi(x_k-x_j)+ \nonu 
\phi(x_j-x_\ell)\phi(x_\ell-x_k)
\biggr] -[x\leftrightarrow y]\: , 
\end{eqnarray*}
and using now the relation in \Ref{rel} and $f(-x)=f(x)$ we get
\begin{eqnarray*}
W_3 =  \sum_{j<k<\ell}(-\lambda^2)\biggl[
f(x_k-x_j) + f(x_j-x_\ell) + f(x_\ell-x_k) \biggr]
 -[x\leftrightarrow y] \nonu 
=  \sum_j \sum_{k\neq j}\sum_{\ell\neq j,k}
\biggl[-\lambda^2 f(x_j-x_k) \biggr] -[x\leftrightarrow y]\nonu
=  -(N-2)\lambda^2\sum_j \sum_{k\neq j}
\biggl[f(x_j-x_k)- f(y_j-y_k)\biggr] \: . 
\end{eqnarray*}
Finally, 
\begin{eqnarray*}
W_4 =   \sum_{j,k} \sum_{\ell \neq k}(-\lambda^2)
\biggl[ \phi(y_k-x_j)\phi(x_j-y_\ell)
+ \phi(y_\ell-y_k)\phi(y_k-x_j)+ \nonu
\phi(y_\ell-y_k)\phi(x_j-y_\ell)\biggr] 
-[x\leftrightarrow y] 
\end{eqnarray*}
where we wrote the same term in two different ways by renaming
summation indices.  We can now use the
relation in \Ref{rel} again, and we obtain
\begin{eqnarray*}
W_4 =   \sum_{j,k} \sum_{\ell \neq k}(-\lambda^2)\biggr[ 
f(y_k-x_j) + f(x_j-y_\ell) + f(y_\ell-y_k) 
\biggr] 
-[x\leftrightarrow y] 
\nonu 
N\lambda^2  \sum_j \sum_{k \neq j} 
\biggr[f(x_j-x_k)-  f(y_j-y_k)  \biggl] 
\end{eqnarray*} 
where the terms even under $[x\leftrightarrow y]$ canceled. Collecting
all terms and using $\phi'(x)=-V(x)$ and $2f(x)=V(x)-\phi(x)^2 + c_0$
we get
\begin{eqnarray*}
W =  \sum_j \sum_{k \neq j}\biggr[ 
\lambda\phi'(x_j-x_k) +\lambda^2\phi(x_j-x_k)^2 -\lambda^2(N-2)f(x_j-x_k)
\nonu +
\lambda^2 N f(x_j-x_k)\biggl] 
- [x\leftrightarrow y] = \nonu
 \sum_j \sum_{k \neq j}\biggr[ 
\lambda\phi'(x_j-x_k) +\lambda^2\phi(x_j-x_k)^2 +
2\lambda^2 f(x_j-x_k)\biggl] - [x\leftrightarrow y] = \nonu
 \sum_j \sum_{k \neq j}\biggr[ \lambda(\lambda-1) V(x_j-x_k) 
-  \lambda(\lambda-1) V(y_j-y_k) \biggr] \: .  
\end{eqnarray*}
We thus see that $WF$ is equal to the r.h.s.\ of \Ref{rem}.
\QED

\appsection{Proof of Proposition~\ref{prop1}}
\label{appB}
We first observe that Lemma~\ref{fact1} remains true if we replace
$F(\vx;\vy)$ by
\begin{equation}
\label{Fp}
F'(\vx;\vy) = c \, \ee{ \ii P \sum_{j=1}^N (x_j-y_j)} \, F(\vx;\vy)
\end{equation}
for arbitrary constants $P\in\R$ and $c\in\C$. [To see this, introduce
center-of-mass coordinates $X=\sum_{j=1}^N x_j/N$ and $x'_j=(x_j-x_1)$
for $j=2,\ldots,N$, and similarly for the $y$'s. Then $H(\vx)= -
\partial^2/\partial X^2 + H_c(\vx')$, and similarly for
$H(\vy)$. Invariance of \Ref{rem} under $F\to \exp{[-\ii P(X-Y)N]}F$
thus follows from $(\partial/\partial X + \partial/\partial
Y)F(\vx;\vy)=0$, and the latter is implied by the obvious invariance
of $F(\vx;\vy)$ under $x_j\to x_j+a$, $y_j\to y_j+a$, $a\in\R$.
Invariance of \Ref{rem} under $F\to c F$ is trivial, of course].

As shown below, one can find constants $c$ and $P$ such that the
function $F'(\vx;\vy)$ above is the generating function for the $\hat
F_\vn(\vx)$ in \Ref{Fhat} as follows,
\begin{equation}
\label{genfun} 
F'(\vx; \vy) = \sum_{\vn\in\Z} \hat
F_\vn(\vx)\xi_1^{-n^+_1}\xi_2^{-n^+_2}\cdots \xi_N^{-n^+_N}\: ,\quad
\xi_j=\ee{\ii y_j}
\end{equation}
with 
\begin{equation}
n^+_j = n_j + \frac{\lambda}2(N+1-2j) \: , 
\end{equation} 
and the series converges absolutely in the following region,
\begin{equation}
\label{region}
1 <|\xi_1|<|\xi_2|<\cdots<|\xi_N|< q^{-2} \: .
\end{equation}
Moreover, in the same region one can change variables in the eCS
Hamiltonian $H(\vy)$ from $y_j$ to $\xi_j$ and expand
\begin{eqnarray}
\label{Hyseries} 
 H(\vy) = \sum_j \left( \xi_j \frac{\partial}{\partial\xi_j}\right)^2 -
 \gamma \sum_{j<k}\sum_{\nu\in\Z} S_\nu
 \left(\frac{\xi_j}{\xi_k}\right)^\nu
\end{eqnarray}
with the coefficients $S_\nu$ in \Ref{Snu}.  Inserting this in
$H(\vx)F'(\vx;\vy)=H(\vy)F'(\vx;\vy)$ (implied by Lemma~\Ref{fact1})
and equating the coefficients of $\xi_1^{-n^+_1}\xi_2^{-n^+_2}\cdots
\xi_N^{-n^+_N}$ on both sides we obtains the identity in \Ref{HFhat}
with $\cE_0(\vn)=\sum_j (n^+_j)^2$ as in \Ref{Enn}.

To prove \Ref{genfun} we note that the definition of the functions
$f_\vn(\vz)$ in \Ref{fn} is equivalent to the following generating
function,
\begin{equation}
\label{fn1}
\frac{\prod_{1\leq j<k\leq N}\Theta(\xi_j/\xi_k)}{
\prod_{j,k=1}^N\Theta(z_j/\xi_k)} = \sum_{\vn\in\Z}
f_\vn(\vz)\xi_1^{-n_1}\xi_2^{-n_2}\cdots \xi_N^{-n_N}\: .
\end{equation}
Moreover, using 
\begin{eqnarray*}
\tet(y) = \half \ee{\ii\pi/2 } \xi^{-1/2}\Theta(\xi)\quad \;
\; \mbox{ for }\; \xi=\ee{\ii y}\; \mbox{ if }\; |\xi|<1
\end{eqnarray*} 
following from \Ref{tet} and the obvious identity $\sin(y/2)=\frac12
\ee{\ii\pi/2} \ee{-\ii y/2}(1-\ee{\ii y})$, we find by a
straightforward computation that $F(\vx;\vy)$ in \Ref{F} is equal to
the expression on the l.h.s.\ of \Ref{fn1} multiplied by $\Psi_0(\vx)$
in \Ref{Psi0} and the factor
\begin{eqnarray*} 
\ee{\ii\pi \lambda[N(N-1)/2-N^2]/2} \frac{ \prod_{j<k}
\ee{-\ii\lambda( y_{j}-y_k )/2} } {\prod_{j,k} \ee{-\ii\lambda(x_j-y_k
)/2 }} = const.\ \ee{ \ii\lambda N \sum_{j=1}^N (x_j-y_j)/2} \ee{
-\ii\lambda \sum_{j=1}^N (N+1-2j)y_j/2} \: .
\end{eqnarray*}
This shows that \Ref{genfun} holds true for certain constants $P$ and
$c$. We are left to prove 
\begin{equation}
\label{Veps} 
V(y) = -\sum_{\nu\in\Z} S_\nu \xi^{-\nu}\; \mbox{ for $\xi=\ee{\ii y}$
such that $|\xi| < 1$}
\end{equation} 
which obviously implies \Ref{Hyseries}. For that we insert the
identity 
\begin{eqnarray*} 
\frac{1}{4\sin^2[(\varphi \pm \ii\eps)/2]} = -\sum_{\nu=1}^\infty \nu
\ee{\pm \ii\nu \varphi-\nu\eps}\; \mbox{for all real $\varphi$ and
$\eps>0$}
\end{eqnarray*} 
in \Ref{V0} and obtain $V(y) = -\sum_{\nu=1}^\infty \nu \left[
\xi^{-\nu} + \sum_{m=1}^\infty \ee{-m\beta}( \xi^\nu + \xi^{-\nu})
\right]$. Summing up the geometric series in $m$ we obtain
\Ref{Veps}.

\appsection{Further proofs} 
\label{appC}
\subsection{Proof of Lemma~\ref{lemPn}}
\label{cPbound} 
We first prove the estimates in \Ref{Thetabound} using the definition
in \Ref{Theta}:
\begin{eqnarray*} 
|\Theta(z)|\leq (1+|z|) \prod_{m=1}^\infty\left[ \left(1+q^{2m}
  |z| \right)\left(1+q^{2m}/|z| \right) \right] =
\Theta(-|z|)\: ,
\end{eqnarray*}
and
\begin{eqnarray*}
|\Theta(z)| \geq (1-|z|) \prod_{m=1}^\infty\left[ \left(1-q^{2m}
|z| \right)\left(1-q^{2m}/|z| \right) \right] = \Theta(|z|)
\end{eqnarray*}
for $q^2\leq|z|\leq 1$. With that we can estimate the integrand in
\Ref{fn} as follows,
\begin{eqnarray*} 
\left| \left( \prod_{j} \xi_j^{n_j} \right)
\frac{\prod_{j<k}\Theta(\xi_j/\xi_k)^\lambda }{\prod_{j,k}\Theta(\ee{\ii
x_j}/\xi_k)^\lambda } \right| \leq \left( \prod_{j} |\xi_j^{n_j}| \right)
\frac{\prod_{j<k}\Theta(-|\xi_j/\xi_j|)^\lambda}{\prod_{j,k}
\Theta(1/|\xi_k|)^\lambda } 
\end{eqnarray*}
where the r.h.s.\ is constant on the integration paths in \Ref{cCj}.
Thus \Ref{fn} implies the following upper bound,
\begin{equation} 
\label{cPbnd} 
|f_\vn(\vz)|\leq \ee{\sum_j \eps_j n_j}
\frac{\prod_{j<k} \Theta(-\ee{-(\eps_k-\eps_j) }
  )^{\lambda}}{\prod_{j,k} \Theta(\ee{-\eps_k} )^{\lambda} } 
\end{equation}
for \textit{any} choice of allowed parameters $\eps_j$. To make the bound
in \Ref{cPbnd} more specific we choose
\begin{eqnarray*} 
\eps_j = \eps( j - \rho\sign(n_j) ),\quad
\eps=\frac{\beta}{N+b},\quad \rho= \frac{b}{1+2b}
\end{eqnarray*} 
with $b>0$ arbitrary; note that these parameters satisfy the
conditions in \Ref{cCj}.  We also observe that
\begin{eqnarray*}
\frac12 \Theta(-x)<\frac{\Theta(-x)}{1+x} < \frac{\Theta(-q^2)}{1+q^2}
<\Theta(-q^2) \; \mbox{ for $q^2<x<1$}
\end{eqnarray*}
and
\begin{eqnarray*}
\frac{\Theta(x)}{1-x_1} \geq \frac{\Theta(x)}{1-x} \geq
\frac{\Theta(x_0)}{1-x_0} > 0 \; \mbox{ for $q^2<x_0\leq x\leq x_1
<1$}\: ;
\end{eqnarray*} 
these estimates are easily proved using $\Theta(\pm x)/(1\mp x)=
\prod_{n=1}^\infty[1+q^{4n}\mp q^{2n}(x+1/x)]$ and the fact that the
function $x+1/x$ is monotonically decreasing in the interval $0<x
<1$.  Thus
\begin{eqnarray*} 
|f_\vn(\vz)| < \ee{\sum_j \eps(j n_j - \rho |n_j|)}\,
\frac{[2\Theta(-q^2)]^{N(N-1)\lambda/2}}{[(1-x_1)\Theta(x_0)/(1-x_0)]^{N^2\lambda}}
\end{eqnarray*}
with $x_0=\min_{k,\vn} \ee{-\eps_k} = \ee{-\eps (N + \rho)}$ and
$x_1=\max_{k,\vn} \ee{-\eps_k} = \ee{-\eps (1 - \rho )}$. Inserting
$\ee{-\eps}=q^{K}$ and $\rho K=\tilde{K}$ we obtain $x_0=q^{2-2b\tilde{K}}$ and
$x_1=q^{K-\tilde{K}}$. \QED

\subsection{Proof of the estimates in \Ref{KsN1}}
\label{KsNestimate}
In this appendix we derive the estimate \Ref{KsN1} for the sum
$K_s(\vm)$ defined in \Ref{KKsN1}.

Representing the Kronecker delta by integrals,
\begin{eqnarray*} 
\delta(\vm ,\mbox{$\sum_{\kappa =1}^s$}\vnu_\kappa ) = \left(
 \prod_{j=1}^N \int_{-\pi}^{\pi}\frac{dy_j}{2\pi} \right) \ee{-\ii
 \vy\cdot(\vm - \sum_\kappa \vnu_\kappa)}
\end{eqnarray*}
and recalling \Ref{vnu}, $\vy\cdot\vnu_\kappa =
\nu_\kappa(y_{j_\kappa}-y_{k_\kappa})$, we obtain
\begin{equation} 
\label{Ksm1} 
K_s(\vm) = |\gamma|^s \left( \prod_{j=1}^N
\int_{-\pi}^{\pi}\frac{dy_j}{2\pi} \ee{-(\ii y_j+ \eps_j) m_j }\right)
\left( \sum_{ 1\leq j<k \leq N}\sum_{\nu\in\Z} S_\nu \ee{\ii
\nu(y_j-y_k) -\nu(\eps_k-\eps_j) } \right)^s
\end{equation}
for $0<\eps_1<\eps_2<\cdots<\eps_N<\beta$
in the limit $\eps_N\downarrow 0$. The shifts in the integration paths,
$y_j\to y_j - \ii \eps_j$, are to guarantee the absolute convergence of
the sums in the integrand.  It it important to note that,
actually, this limit need not be taken: changing variables to
$\xi_j=\ee{\ii y_j + \eps_j}$ the latter integral can be written as
follows,
\begin{eqnarray*}
K_{s}(\vm) = |\gamma|^s \left( \prod_{j=1}^N \oint_{\cC_j}\frac{\dd
\xi_j}{2\pi\ii \xi_j} \xi_j^{-m_j} \right) \left( \sum_{ j<k
}\sum_{\nu\in\Z} S_\nu \left(\frac{\xi_j}{\xi_k}\right)^\nu
\right)^{s}
\end{eqnarray*} 
with the integration paths $\cC_j$ defined in \Ref{cCj}.  The
integrand is analytic for all $\eps_j$ obeying the condition in
\Ref{cCj}, as can be checked by recalling the definition of $S_\nu$ in
\Ref{Snu} and $q^2=\ee{-\beta}$. Thus Cauchy's theorem implies that
integral on the r.h.s.\ in \Ref{Ksm1} is independent of the $\eps_j$.

Inserting $\eps_j=j\eps$ where $0<\eps<\beta/N$ and using the triangle
inequality and $S_\nu\geq 0$ we deduce from \Ref{Ksm1} that 
\begin{eqnarray} 
|K_s(\vm)| \leq |\gamma|^s \left( \prod_{j=1}^N
\int_{-\pi}^{\pi}\frac{dy_j}{2\pi} |\ee{-(\ii y_j  + j \eps) m_j
}|\right) \left( \sum_{j<k}\sum_{\nu\in\Z} S_\nu |\ee{\ii \nu(y_j-y_k)
  -\nu (k-j)\eps )}| \right)^s \nonu = \ee{-\sum_j j \eps m_j } \left(
|\gamma| \sum_{ j<k }\sum_{\nu\in\Z} S_\nu \ee{-\nu (k-j)\eps}
\right)^{s} \: . 
\end{eqnarray}
To simplify this bound we use $S_\nu \leq |\nu| q^{|\nu|-\nu}/(1-q^2)$
following from the definition in \Ref{Snu}, and thus
\begin{eqnarray*} 
\sum_{ j<k } \sum_{\nu\in\Z} S_\nu \ee{-\nu (k-j)\eps} \leq 
\sum_{j<k} \sum_{\nu\in\Z} |\nu| \frac{q^{|\nu|-\nu}
}{(1-q^2)} \ee{-\nu(k-j)\eps}=  \nonu \frac{1}{(1-q^2)} \sum_{j<k}
\sum_{\nu=1}^\infty \nu \left( \ee{-\nu(k-j)\eps} + \ee{-\nu[\beta
-(k-j)\eps]} \right)=  \nonu \frac{1}{(1-q^2)} \sum_{\ell=1}^{N-1}
\sum_{\nu=1}^\infty \nu\left( (N-\ell) \ee{-\nu\ell\eps} + \ell
\ee{-\nu[\beta -(N-\ell)\eps]} \right) = \nonu 
\frac{1}{(1-q^2)}\sum_{\ell=1}^{N-1} \left( (N-\ell)
\frac{\ee{-\ell\eps}}{(1-\ee{-\ell \eps})^2} + \ell \frac{\ee{-[\beta
-N\eps]-\ell\eps} }{(1 - \ee{-[\beta -N\eps]-\ell\eps})^2 } \right)
\end{eqnarray*}
where we used $q^2=\ee{-\beta}$ and changed summation
variables. Inserting the estimate
\begin{eqnarray*}
\frac{p}{(1-px)^2} < \frac1{(1-x)^2}\quad \; \mbox{ if $0<p<1$ and
  $0<px^2<1$}
\end{eqnarray*} 
for $p=\exp(-[\beta-N\eps])$ and $x=\exp(-\eps)$ we get
\begin{eqnarray*} 
\sum_{ j<k }\sum_{\nu\in\Z} S_\nu \ee{-\nu (k-j)} 
< \frac{N}{(1-q^2)}\sum_{\ell=1}^{N-1}
\frac{\ee{-\ell\eps}}{(1-\ee{-\ell \eps})^2}< \nonu
\frac{
N}{(1-q^2)}\sum_{\ell=1}^{N-1}
\frac{\ee{-\ell\eps}}{(1-\ee{-\eps})^2} < \frac{N(N-1) \ee{-\eps}
}{(1-\ee{-\eps})^3} \: ,
\end{eqnarray*}
since $(1-\ee{-(N-1)\eps})<(1-q^2)$. To summarize, 
\begin{equation} 
\label{generalKsbound}
K_s(\vm) < \ee{-\sum_j j \eps  m_j } \left( \frac{|\gamma| N(N-1) 
  \ee{-\eps} }{(1-\ee{-\eps})^3}\right)^s\; \mbox{ if }\;  0<\eps <
\frac{\beta}{N}\: .
\end{equation}
Setting $\eps=\beta/(N+b)$ for $b>0$ we get the bounds in
\Ref{KsN1}. \QED

\begin{remark}
\label{remC1} 
It is instructive to see in more details how the upper bounds for
$K_s(\vzero)$ come about: It is easy to see that the contributions of
lowest order in $q^2$ come from the identity in \Ref{cycle}, for
example for $s\leq N$ the lowest order contributions are for
$\vnu_r=-\vE_{j_rj_{r+1}}$ for $r=1,2,\ldots,s-1$ and
$\vnu_s=\vE_{j_1j_s}$. Thus for $s\leq N$,
$K_s(\vzero)=c_{N,s}|\gamma|^s(S_{-1})^{s-1}S_1+\ldots = c_{N,s}
|\gamma|^s q^2+\ldots$ with the dots indicating higher order terms,
and, more generally, $K_s(\vzero) = c_{N,s} |\gamma|^s
q^{2\lceil\frac{s}{N}\rceil} + \cO(q^{2\lceil\frac{s}{N}\rceil+2})$;
the constants $c_{N,s}$ are combinatorial factors. This and
\Ref{generalKsbound} for $\eps=\beta/N$ and $\vm=\vzero$ suggest the
improved estimate in \Ref{ConjKsbound}.

\end{remark}

\end{document}